\documentclass[aps,prl,twocolumn]{revtex4-1}
\usepackage[T1]{fontenc}
\usepackage[latin1]{inputenc}
\usepackage{lmodern}
\expandafter\let\csname equation*\endcsname\relax
\expandafter\let\csname endequation*\endcsname\relax
\usepackage{amsmath}
\usepackage{esint}
\usepackage{amssymb}
\usepackage{graphicx}
\usepackage{xcolor}
\usepackage{natbib}
\usepackage{bm}
\usepackage{bbold}
\usepackage[mathscr]{euscript}
\usepackage[cal=boondoxo]{mathalfa}
\usepackage{comment}

\def\pM{\mathrel{\raise 2pt \hbox{\tiny(}\!\raise 1pt \hbox{+}\settowidth {\dimen03} {+}\hskip-\dimen03 \raise -2.4pt \hbox {$-$} \!\raise 2pt \hbox{\tiny)}}}

\begin{document}
\title{Strain-induced excitonic instability in twisted bilayer graphene}
\author{H\'ector Ochoa}
\affiliation{Department of Physics, Columbia University, New York, NY 10027, USA}

%\date{\today}
\begin{abstract}
The low-energy bands of twisted bilayer graphene form Dirac cones with approximate electron-hole symmetry at small rotation angles. These crossings are protected by the emergent symmetries of moir\'e patterns, conferring a topological character to the bands. Strain accumulated between layers (\textit{heterostrain}) shifts the Dirac points both in energy and momentum. The overlap of conduction and valence bands favors an excitonic instability of the Fermi surface close to the neutrality point. The spontaneous condensation of electron-hole pairs breaks time reversal symmetry and the separate conservation of charge within each valley sector. The order parameter describes interlayer circulating currents in a Kekul\'e-like orbital magnetization density wave. 
Vortices in this order parameter carry fermion numbers owing to the underlying topology of the bands. This mechanism may explain the occurrence of insulating states at neutrality in the most homogenous samples, where uniform strain fields contribute both to stabilizing the relative orientation between layers and to the formation of an excitonic gap.
\end{abstract}
\maketitle

When two coupled graphene layers are rotated with respect to each other the electronic spectrum is reorganized in narrow bands characterized by a diminished Fermi velocity \cite{portu1,Andrei}, which cancels at a relative twist of about $\theta\approx 1.1^{\textrm{o}}$ \cite{Morell_etal,MacDonald}. Devices around this \textit{magic angle} are insulating when an approximately integer number of electrons per moir\'e supercell is either added or removed from the system at low temperatures \cite{Jarillo1}. The system behaves as a superconductor when some of the insulators are doped \cite{Jarillo2}. This phenomenology has been reproduced and extended \cite{Columbia,Stanford,ICFO,SantaBarbara,screening1,screening2,screening3,screening4}, but there are important differences from device to device. 

This work targets the situation around the charge neutrality point. Band models predict a Dirac semimetal, reminiscence of the original Dirac cones on each individual layer protected by the emergent symmetries of the moir\'e pattern \cite{Boston}. Semimetallic behavior is observed in most transport devices, with the exceptions of Refs.~\onlinecite{ICFO}~and~\onlinecite{SantaBarbara}. In the latter case, this behavior can be attributed to the alignment with one of the encapsulating boron nitride layers. However, the samples of Ref.~\onlinecite{ICFO} were very homogenous, resulting in enlarged superconducting domes, which suggests that the insulating behavior observed at neutrality was also a many-body effect. Signatures of this many-body gap are also observed in scanning tunneling microscopy (STM) studies \cite{STM1,STM2,STM3,STM4}.

\begin{figure}
\centerline{\includegraphics[width=\linewidth]{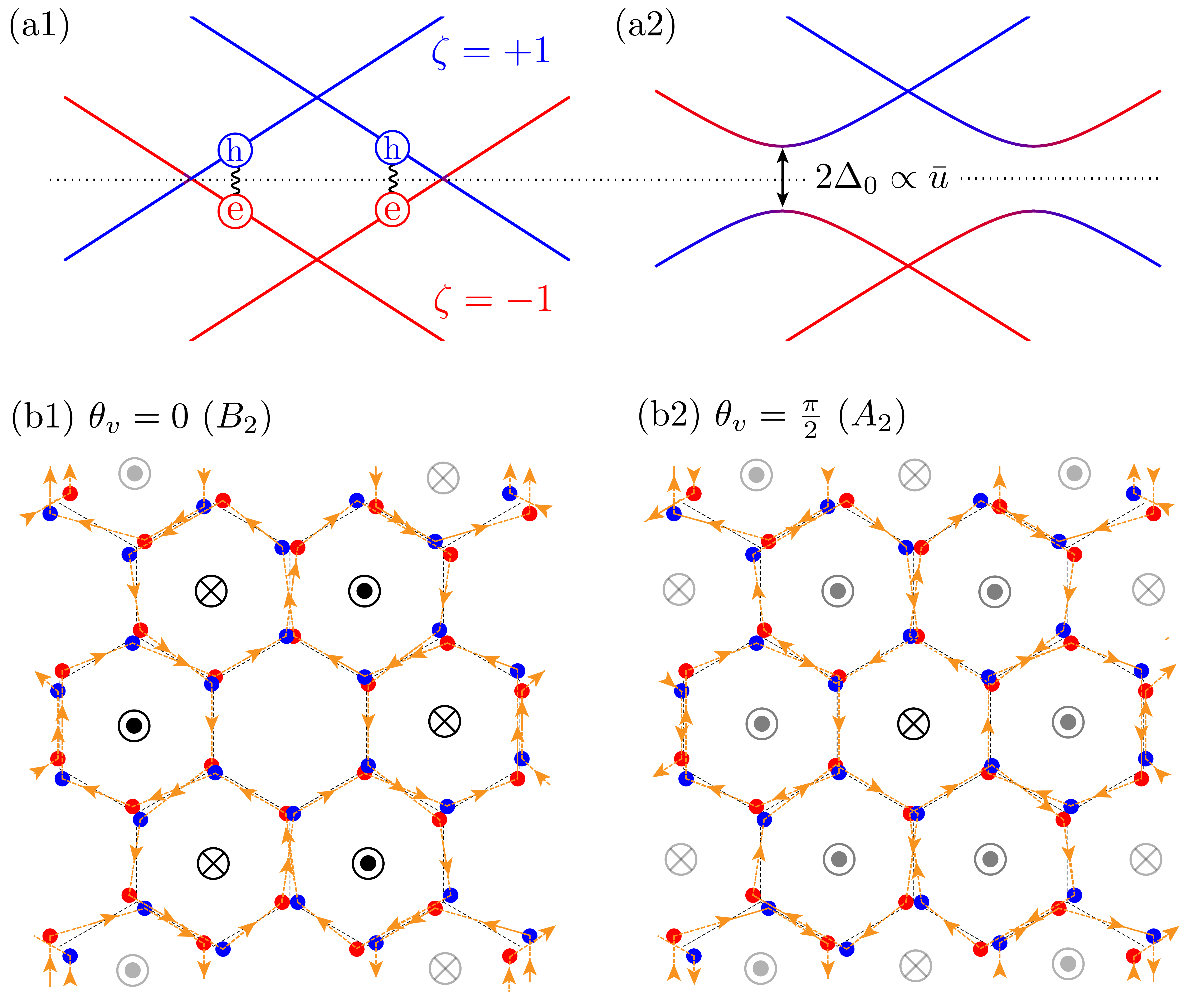}}
\caption{(a) Heterostrain shifts the Dirac points in energy. The valleys are inverted in time-reversed points of reciprocal space. Electron-hole band nesting makes the Fermi surface unstable. The resulting energy gap is determined by the amplitude of the excitonic condensate, $\Delta_0$. (b) Kekul\'e-like patterns of circulating currents corresponding to the two components of the excitonic order parameter, Eq.~\eqref{eq:order_parameter}. $B_2$ and $A_2$ label the corresponding symmetry representation (see Table~\ref{tab:operators}). Currents circulate between the top and bottom layers, represented by blue and red dots, respectively. This circulation gives rise to orbital magnetization density waves represented by the symbols at the center of the hexagonal plaquettes.}
\label{fig:fig1}
\end{figure}

\begin{table*}
\centering
\begin{tabular}{|c||c|c|c|c|c|c|}
\hline
valley& $A_1$ & $A_2$ & $B_1$ & $B_2$ & $E_1$ & $E_2$   \\
\hline
\hline
0 & $\hat{1}$ (+) & $\hat{\Sigma}_z$ (-) & $\hat{\Sigma}_z\hat{\Gamma}_z$ (+) & $\hat{\Gamma}_z$ (-) & $\left[\begin{array}{c} \hat{\Sigma}_x\\ \hat{\Sigma}_y\end{array}\right]$ (-) & $\left[\begin{array}{c} -\hat{\Sigma}_y\hat{\Gamma}_z\\ \hat{\Sigma}_x\hat{\Gamma}_z\end{array}\right]$ (+) \\
\hline
$x$ & $\hat{\Sigma}_z\hat{\Lambda}_x$ (+) & $\hat{\Lambda}_x$ (-) & $\hat{\Lambda}_x\hat{\Gamma}_z$ (+) & $\hat{\Sigma}_z\hat{\Lambda}_x\hat{\Gamma}_z$ (-) & $\left[\begin{array}{c} -\hat{\Sigma}_y\hat{\Lambda}_x\\ \hat{\Sigma}_x\hat{\Lambda}_x\end{array}\right]$ (+) &  $\left[\begin{array}{c} \hat{\Sigma}_x\hat{\Lambda}_x\hat{\Gamma}_z\\ \hat{\Sigma}_y\hat{\Lambda}_x\hat{\Gamma}_z\end{array}\right]$ (-) \\
\hline
$y$ & $\hat{\Lambda}_y\hat{\Gamma}_z$ (+) & $\hat{\Sigma}_z\hat{\Lambda}_y\hat{\Gamma}_z$ (-) & $\hat{\Sigma}_z\hat{\Lambda}_y$ (+) & $\hat{\Lambda}_y$ (-) & $\left[\begin{array}{c} \hat{\Sigma}_x\hat{\Lambda}_y\hat{\Gamma}_z\\ \hat{\Sigma}_y\hat{\Lambda}_y\hat{\Gamma}_z\end{array}\right]$ (-) &  $\left[\begin{array}{c} -\hat{\Sigma}_y\hat{\Lambda}_y\\ \hat{\Sigma}_x\hat{\Lambda}_y\end{array}\right]$ (+)  \\
\hline
$z$ & $\hat{\Sigma}_z\hat{\Lambda}_z\hat{\Gamma}_z$ (-) & $\hat{\Lambda}_z\hat{\Gamma}_z$ (+) & $\hat{\Lambda}_z$ (-) & $\hat{\Sigma}_z\hat{\Lambda}_z$ (+) & $\left[\begin{array}{c} -\hat{\Sigma}_y\hat{\Lambda}_z\hat{\Gamma}_z\\ \hat{\Sigma}_x\hat{\Lambda}_z\hat{\Gamma}_z\end{array}\right]$ (-) & $\left[\begin{array}{c} \hat{\Sigma}_x\hat{\Lambda}_z\\ \hat{\Sigma}_y\hat{\Lambda}_z\end{array}\right]$ (+)  \\
\hline
\end{tabular}
\caption{Diagonal operators in \textit{mini-valley} in the basis of Bloch waves~\eqref{eq:basis}. The first row indicates the corresponding irreducible representation of $D_6$. The first column labels singlet (0) and triplet ($x,y,z$) representations of valley rotations. The sign between brackets indicates the parity under time-reversal symmetry, $\mathcal{T}$.}
\label{tab:operators}
\end{table*}

Here I show that layer-asymmetric strain fields favor the formation of an excitonic gap. The proposed scenario is closely related to the incommensurability of the samples at small twist angles. The most generic form of disorder consists of spatially modulated strains accumulated between the layers (heterostrain), usually manifested as inhomogeneities of the beating pattern \cite{STM1,STM2,STM3,STM4}. These inhomogeneities occur because the samples are only metastable, where strong fluctuations in twist angle arise from soft collective modes describing the sliding motion of one layer with respect to the other \cite{phasons}. Strains generated during the fabrication process, on the other hand, can contribute to freezing these modes. Hence, homogenous samples are likely to be the result of uniform strain fields that stabilize their relative alignment. Simultaneously, these strain fields modify the electronic spectrum \cite{strain1,strain2,Fu} and favor specific forms of symmetry breaking among the variety of almost degenerate insulating states predicted by, e.g., Hartree-Fock theory \cite{MacDonald2,Vishwanath1,Vishwanath2,Liu_Dai,Zhang_etal,Gonzalez_Stauber,Cea_Guinea}. The basic idea is illustrated in Fig.~\ref{fig:fig1}(a). Uniform heterostrain fields shift the Dirac points both in quasi-momentum and energy \cite{phasons,Fu} while preserving the electron-hole symmetry of the spectrum at small twist angles \cite{Koshino}. The resulting Fermi surface is unstable with respect to the condensation of electron-hole pairs \cite{Keldysh_Kopaev}. The condensate breaks time reversal $\mathcal{T}$ and U$_{v}$(1) valley symmetries. Depending on the electron-hole pairing phase, the order parameter describes different patterns of circulating currents represented in Fig.~\ref{fig:fig1}(b).

\begin{figure}
\centerline{\includegraphics[width=\linewidth]{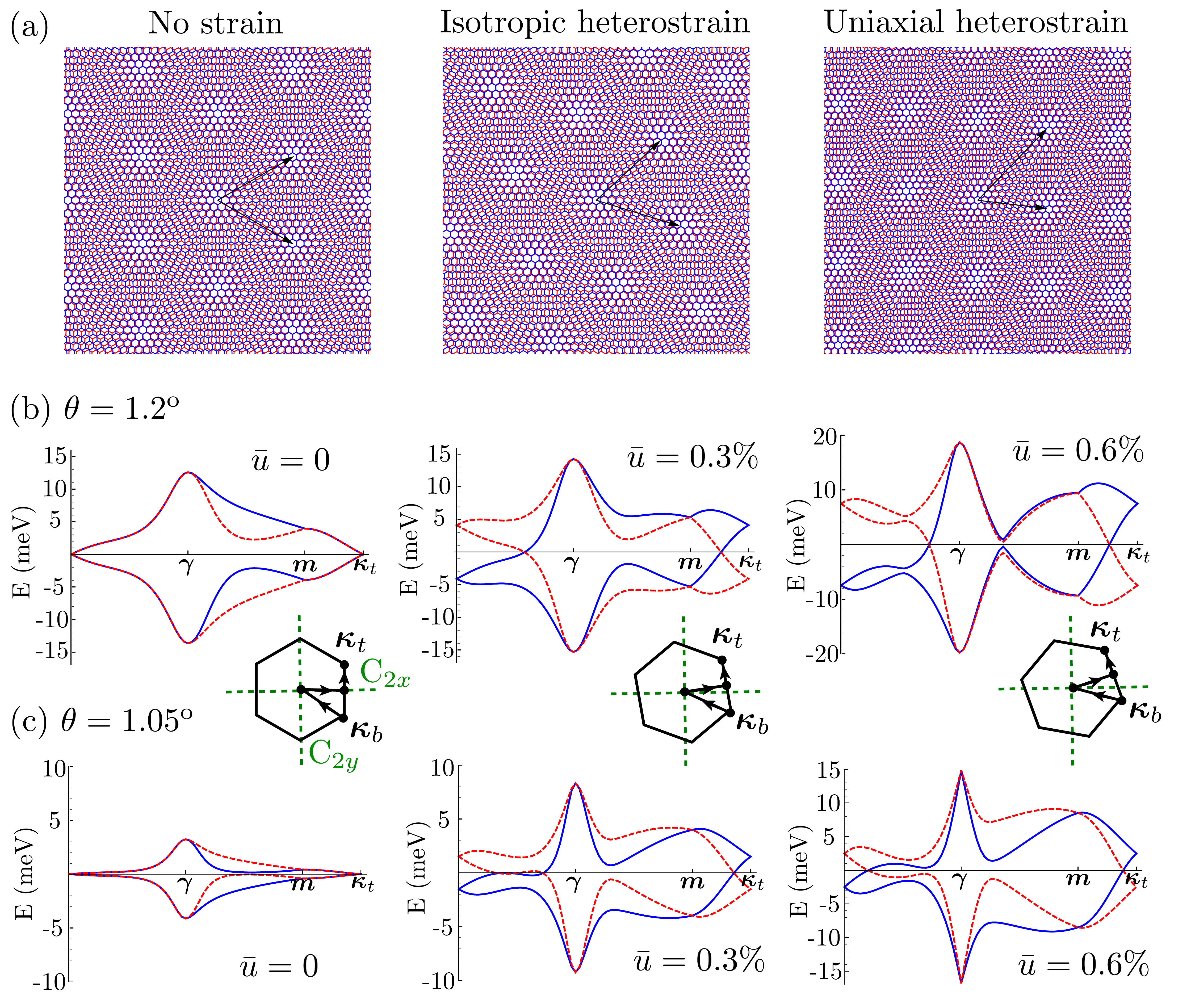}}
\caption{(a) Moir\'e patterns with no strain (left) and isotropic heterostrain (right); the twist angle is the same in both images. Heterostrain tilts the beating pattern with respect to the orientation of the atomic lattices. (b) Lowest-energy bands above the magic angle ($\theta=1.2^{\textrm{o}}$) and different values of isotropic heterostrain. The insets show the path in the corresponding moir\'e Brillouin zone. (c) The same for the nominal magic angle, $\theta=1.05^{\textrm{o}}$. The bands in (b) and (c) were obtained from the continuum model \cite{portu1,MacDonald} with the parameters of Ref.~\onlinecite{Koshino_Fu} and deformation potential constant $D=10$ eV.}
\label{fig:fig2}
\end{figure}

If there is no strain in the system, the relative twist $\theta$ defines a moir\'e pattern of pitch $L_m=a/2\sin(\theta/2)$, where $a$ is graphene's lattice constant; for future reference, $x$ coordinate is defined along a halfway direction between zig-zag axes of the two layers. For small $\theta$, the approximate translational symmetry folds back the position of the microscopic valleys ($\zeta=\pm 1$ for $\mathbf{K}_{\pm}$) onto the two inequivalent corners of the moir\'e Brillouin zone, $\mathbf{K}_{\pm}^{(\nu)}\cong\pm\boldsymbol{\kappa}_{\nu}$, where $\nu=t,b$ labels the two layers and the corresponding points in momentum space (see insets in Fig.~\ref{fig:fig2}). The low-energy properties of the system are assumed to be dominated by charge excitations around a pair of Dirac crossings on each $\boldsymbol{\kappa}_{\nu}$ point, described by the fermionic action $S[\boldsymbol{\hat{\psi}},\boldsymbol{\hat{\psi}}^{\dagger}]=\int_0^{\hbar\beta} d\tau\,\{\int d\mathbf{r}\,\boldsymbol{\hat{\psi}}^{\dagger}(\hbar\partial_{\tau}-\mu)\boldsymbol{\hat{\psi}}+H[\boldsymbol{\hat{\psi}},\boldsymbol{\hat{\psi}}^{\dagger}]\}$, where $\boldsymbol{\hat{\psi}}(\tau,\mathbf{r})$ is a smoothly varying 8-component field and $\mu$ represents the chemical potential. The relation between $\boldsymbol{\hat{\psi}}(\tau,\mathbf{r})$ and the microscopic field operator reads as $\hat{\Psi}(\tau,\mathbf{r})=\boldsymbol{u}(\mathbf{r})\cdot\boldsymbol{\hat{\psi}}(\tau,\mathbf{r})$, where $\boldsymbol{u}\left(\mathbf{r}\right)$ is a vector formed by the Bloch wave functions $u_{\lambda,\zeta,\boldsymbol{\kappa}_{\nu}}(\mathbf{r})$ of the Dirac points,
\begin{align}
\label{eq:basis}
\boldsymbol{u}\left(\mathbf{r}\right)=\left[\vec{u}_{+,\boldsymbol{\kappa}_t}\left(\mathbf{r}\right),\vec{u}_{-,\boldsymbol{\kappa}_t}\left(\mathbf{r}\right),\vec{u}_{+,\boldsymbol{\kappa}_b}\left(\mathbf{r}\right),\vec{u}_{-,\boldsymbol{\kappa}_b}\left(\mathbf{r}\right)\right]^T.
\end{align}
Here $\vec{u}_{\zeta,\boldsymbol{\kappa}_{\nu}}(\mathbf{r})=[u_{1,\zeta,\boldsymbol{\kappa}_{\nu}}(\mathbf{r}),\zeta u_{2,\zeta,\boldsymbol{\kappa}_{\nu}}(\mathbf{r})]^T$ are Dirac spinors of opposite chirality on each valley, where $\lambda=1,2$ labels the complex eigenvalue of the Bloch wave function under C$_{3z}$ rotations \cite{SM}.

The form of the effective Hamiltonian $H=H_0+H_C$ is constrained by the emergent symmetries of long moir\'e patterns protecting the Dirac points: $D_6$ and approximate conservation of charge within each valley sector, U${_c}$(1)$\times$U$_{v}$(1) (here U${_c}$(1) is the global charge symmetry). Operators in the Hilbert space of wave functions~\eqref{eq:basis} can be expressed in a basis of 64 independent hermitian matrices, the identity $\hat{1}$ associated with U$_c$(1), and three inter-commuting Pauli algebras, $\{\hat{\Sigma}_i\}_{\lambda}$, $\{\hat{\Lambda}_i\}_{\zeta}$, $\{\hat{\Gamma}_i\}_{\boldsymbol{\kappa}_{\nu}}$, generating rotations in isospin ($\lambda$), valley and pseudo-spin or \textit{mini-valley} ($\boldsymbol{\kappa}_{\nu}$) indices. Table~\ref{tab:operators} shows all the possible diagonal operators in pseudo-spin space (i.e., invariant under moir\'e translations) classified according to the irreducible representations of $D_6$ \cite{SM}. With this, we can systematically construct all the possible terms in the Hamiltonian and identify the associated broken symmetries. The matrix elements can be estimated from the continuum model \cite{portu1,MacDonald}. In doing so, we must note that as the twist angle decreases the electron-hole symmetry of the original $k\cdot p$ expansion is effectively recovered. This approximate symmetry is implemented by an anti-unitary operator exchanging layers and sublattices \cite{Koshino,SM} which, in combination with time-reversal symmetry $\mathcal{T}$, defines a unitary \textit{chiral} symmetry (different from the one discussed in Ref.~\onlinecite{chiral_model}) relating positive and negative energy eigenstates at the same $\mathbf{q}$ point:\begin{align}
\label{eq:chiral}
\hat{\mathcal{C}}=\hat{\Sigma}_z\hat{\Lambda}_y\hat{\Gamma}_z.
\end{align}

The single-particle term in the Hamiltonian, $H_0[\boldsymbol{\hat{\psi}},\boldsymbol{\hat{\psi}}^{\dagger}]=\int d\mathbf{r}\,\boldsymbol{\hat{\psi}}^{\dagger}\left(\tau,\mathbf{r}\right)\hat{\mathcal{H}}_{0}\,\boldsymbol{\hat{\psi}}\left(\tau,\mathbf{r}\right)$, describes the band dispersion and the coupling with heterostrain, $\hat{\mathcal{H}}_0=\hat{\mathcal{H}}_{\textrm{b}}+\hat{\mathcal{H}}_{\textrm{str}}$. In a series expansion in $\boldsymbol{p}=-i\hbar\boldsymbol{\partial}$, momentum deviations from $\boldsymbol{\kappa}_{\nu}$ points, the dominant terms in the band Hamiltonian are those compatible with the approximate electron-hole symmetry, $\{\hat{\mathcal{C}},\hat{\mathcal{H}}_{\textrm{b}}\}=0$; up to second order in $\boldsymbol{p}$, we have\begin{align}
\label{eq:band_Hamiltonian}
\hat{\mathcal{H}}_{\textrm{b}}=\hbar v_F^*\,\hat{\boldsymbol{\Sigma}}\cdot\boldsymbol{p}+\gamma\left[(p_x^2-p_y^2)\hat{\Sigma}_y\hat{\Gamma}_z+2p_xp_y\hat{\Sigma}_x\hat{\Gamma}_z\right],
\end{align}
where $ v_F^*\approx\frac{1-3\alpha^2}{1+6\alpha^2} v_F$ \cite{MacDonald} and $\gamma\approx\frac{3\alpha}{1+6\alpha^2}\frac{w}{\left|\mathbf{k}_0\right|^2}$. Here $\alpha=w/\hbar v_F|\mathbf{k}_0|$ is the ratio between the interlayer coupling $w$ and the geometrical energy scale defined by the shift in the position of the valleys, $\mathbf{k}_0=\boldsymbol{\kappa}_b-\boldsymbol{\kappa}_t$, and the velocity of graphene Dirac electrons, $v_F$.

\begin{figure}[b!]
\centerline{\includegraphics[width=\linewidth]{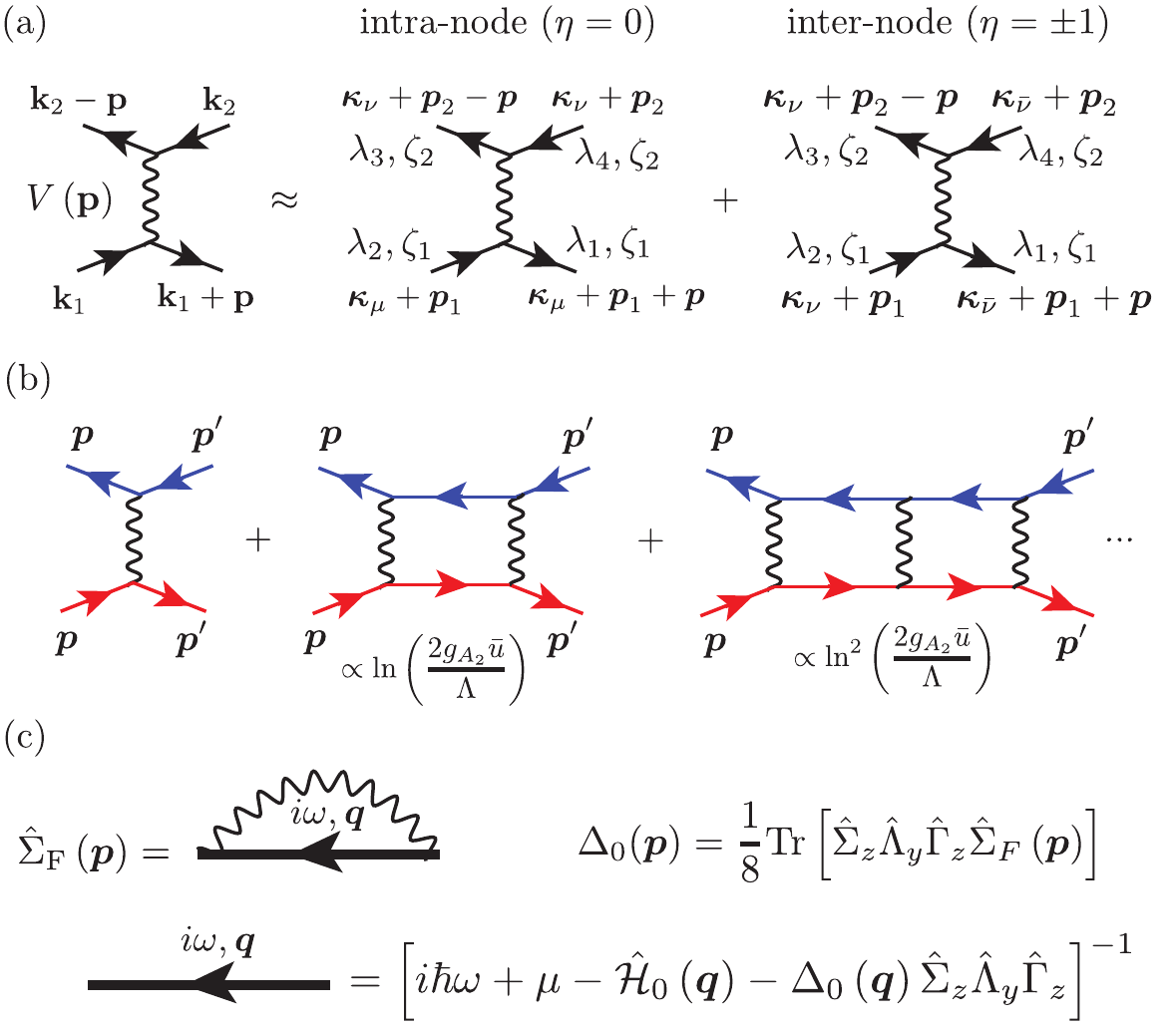}}
\caption{(a) The projected long-range Coulomb interaction consists of \textit{inter-node} ($\eta=0$) and \textit{intra-node} ($\eta=\pm 1$) momentum exchange vertices. (b) The ladder series in the electron-hole channel contain logarithmically divergent contributions resulting from simultaneous resonances in the valley-resolved single-particle Green functions evaluated on the Fermi surface (here $\Lambda$ is an infrared cutoff). (c) Diagramatics of the BCS-like mean field leading to Eq.~\eqref{eq:gap_equation}.}
\label{fig:fig3}
\end{figure}

The strain accumulated between the two layers, $\omega_{ij}=u_{ij}^{t}-u_{ij}^{b}$, couples to electrons as\begin{align}
\nonumber
\hat{\mathcal{H}}_{\textrm{str}}=\, & g_{A_2}\, \omega_{ii}\hat{\Lambda}_z\hat{\Gamma}_z+g_{E_2}^{(1)}\left[\left(\omega_{xx}-\omega_{yy}\right)\hat{\Sigma}_x\hat{\Gamma}_z-2\omega_{xy}\hat{\Sigma}_y\hat{\Gamma}_z\right]\\
& +g_{E_2}^{(2)}\left[2\omega_{xy}\hat{\Sigma}_x\hat{\Lambda}_z+\left(\omega_{xx}-\omega_{yy}\right)\hat{\Sigma}_y\hat{\Lambda}_z\right],
\label{eq:strain}
\end{align}
where $g_i$ are phenomenological couplings allowed by $D_6$ symmetry. There are two main contributions to these: i) Strains on each layer modify the energetics of Dirac electrons through the electron-phonon coupling. ii) The deformation of the beating pattern modifies the moir\'e superlattice potential due to an additional shift of the microscopic valleys \cite{phasons,Fu,SM}. To understand the origin of the latter, consider for a moment the case of isotropic heterostrain, $\omega_{ij}=\bar{u}\,\delta_{ij}$, represented in Fig.~\ref{fig:fig2}(a). The six-fold rotational symmetry of the beating pattern is preserved, but it is tilted with respect to the atomic lattices and angle $\sim \bar{u}/\theta$. In-plane C$_2$ symmetries are broken, and the Dirac cones from the same valley sector and no longer degenerate in energy. This effect is described by the first term in Eq.~\eqref{eq:strain}; perturbation theory gives $g_{A_2}\approx\frac{1-6\alpha^2}{2+12\alpha^2}\,D+\frac{4\pi\alpha^2}{1+6\alpha^2}\, \frac{\hbar v_F}{a}$, where $D\sim10$ eV is the deformation potential constant. As the twist angle decreases, interlayer hybridization $\alpha$ increases, and the geometric contribution (second term) starts to dominate over the electron-phonon coupling. 

Figure~\ref{fig:fig2} shows the lowest energy bands calculated within the continuum model \cite{portu1,MacDonald} for a twist angle above (panel b) and at the nominal magic angle (panel c) for the model parameters in Ref.~\onlinecite{Koshino_Fu}. The Dirac points are shifted in energy several meVs. As noted in Ref.~\onlinecite{Fu}, heterostrain introduces a cutoff for the bandwidth/kinetic energy of electrons, setting a lower bound for the group velocity $v_F^*$. Anisotropic fields break C$_3$ symmetry and displaces the Dirac cones from $\boldsymbol{\kappa}_{\nu}$ points, which is described by the two remaining terms in Eq.~\eqref{eq:strain}. The coupling in the second line is subleading \cite{phasons} so that the electronic spectrum respects the approximate electron-hole symmetry \cite{Fu}. Deviations from the sketch in Fig.~\ref{fig:fig1}(a) do not alter the basic premise: uniform strain fields accumulated between the layers give rise to a semimetallic band structure with overlapping conduction and valence bands. This nested Fermi surface at the neutrality point resembles the cases of AA-stacked bilayer graphene \cite{AA1,AA2} or graphene in the presence of a Zeeman field \cite{KT} with exchanged roles of spin and valley. The large density of states associated with the reduction of the Fermi velocity leads to a broken-symmetry ground state in the presence of electron-electron interactions.

The effective interaction Hamiltonian $H_C$ includes the vertices represented in Fig.~\ref{fig:fig3}(a) \cite{SM}. The \textit{dominant} terms correspond to electron-hole scattering processes like those in Fig.~\ref{fig:fig3}(b), which are large in all orders of perturbation theory due to the band overlap. This marks an instability towards electron-hole pairing, $\Delta_0(\boldsymbol{p})\propto\langle \hat{c}_{+s}^{\dagger}(\boldsymbol{p}) \hat{c}_{-s}(\boldsymbol{p}) \rangle$, where $\hat{c}_{\pm s}^{\dagger}\left(\boldsymbol{p}\right)=\boldsymbol{\hat{\psi}}^{\dagger}_{\boldsymbol{p}}\cdot\boldsymbol{\tilde{u}}_{\pm s}\left(\boldsymbol{p}\right)$ and $\boldsymbol{\tilde{u}}_{\pm s}\left(\boldsymbol{p}\right)$ follows from the diagonalization of $\hat{\mathcal{H}}_0$ with eigenvalue $\varepsilon_{\pm s}(\boldsymbol{p})\approx \pm(\hbar v_F^*|\boldsymbol{p}|-2sg_{A_2}\bar{u})$ and index $s=\pm1$ in the two mini-valleys; the associated Bloch wave functions are $u_{\pm s,\boldsymbol{\kappa}_{\nu}+\boldsymbol{p}}(\mathbf{r})\approx e^{i\boldsymbol{p}\cdot\mathbf{r}}\,\boldsymbol{\tilde{u}}_{\pm s}\left(\boldsymbol{p}\right)\cdot\mathbf{u}(\mathbf{r})$. Neglecting spin for the moment, electron-hole pairing involves correlations between the internal degrees of freedom described by a matrix $\hat{\Delta}(\boldsymbol{p})$ in 8-spinor space, $\Delta[\boldsymbol{\hat{\psi}},\boldsymbol{\hat{\psi}}^{\dagger}]=\sum_{s,\boldsymbol{p}}\Delta_0\left(\boldsymbol{p}\right)\,\hat{c}_{-s}^{\dagger}\left(\boldsymbol{p}\right)\hat{c}_{+s}\left(\boldsymbol{p}\right)+\textrm{h.c.}=\sum_{\boldsymbol{p}}\boldsymbol{\hat{\psi}}_{\boldsymbol{p}}^{\dagger}\,\hat{\Delta}\left(\boldsymbol{p}\right)\boldsymbol{\hat{\psi}}_{\boldsymbol{p}}
$. The \textit{chiral} symmetry in Eq.~\eqref{eq:chiral} imposes some relations in the wave functions, in particular, there is always a gauge in which $\boldsymbol{\tilde{u}}_{\pm s}\left(\boldsymbol{p}\right)= \hat{\mathcal{C}}\,\boldsymbol{\tilde{u}}_{\mp s}\left(\boldsymbol{p}\right)$. Naively, we could just identify $\hat{\Delta}(\boldsymbol{p})=\Delta_0(\boldsymbol{p})\hat{\mathcal{C}}$, however, U$_{v}$(1) rotations do not change the energy of the condensate in the continuum theory. Thus, the order parameter is parametrized by a phase $\theta_v$,\begin{align}
\label{eq:order_parameter}
\hat{\Delta}\left(\boldsymbol{p}\right)=\Delta_0\left(\boldsymbol{p}\right)\,\boldsymbol{\hat{n}}_v\cdot\hat{\Sigma}_z\hat{\boldsymbol{\Lambda}}\hat{\Gamma}_z,\,\,\, \boldsymbol{\hat{n}}_{v}=\left(\cos\theta_v,\sin\theta_v\right).
\end{align}
The mean-field Hamiltonian $\hat{\mathcal{H}}_0+\hat{\Delta}$ gives rise to four branches of charge excitations with a gap determined by the amplitude of the order parameter. The meaning of $\hat{\Delta}$ follows from its matrix structure in the Bloch wave basis~\eqref{eq:basis}. It can be understood as an orbital magnetization density wave with maximum amplitude at AA stacked regions of the moir\'e cell with a Kekul\'e-like modulation on the atomic scale. The exact microscopic profile is determined by the pairing phase $\theta_v$. In the gauge of Eq.~\eqref{eq:basis}, $x$ and $y$ components correspond to the high-symmetric patterns represented in Fig.~\ref{fig:fig1}(b).

The divergent ladder series can be summed up in the BCS-like mean field approach depicted in Fig.~\ref{fig:fig3}(c). Projecting the Fock self-energy to the matrix structure of the order parameter, I obtain 
at neutrality $\mu=0$:\begin{widetext}\begin{align}
\label{eq:gap_equation}
\Delta_0\left(\boldsymbol{p}\right)=\sum_{s=\pm 1}\int\frac{d\boldsymbol{q}}{\left(2\pi\right)^2}\sum_{\eta=0,\pm1}\sum_{\left\{\mathbf{G}\right\}} V\left(\boldsymbol{q}-\boldsymbol{p}+\eta\mathbf{k}_0+\mathbf{G}\right)\frac{f_{s}^{(\eta)}\left(\boldsymbol{q},\boldsymbol{p},\mathbf{G}\right)\,\Delta_0\left(\boldsymbol{q}\right)}{2\sqrt{\varepsilon_{\pm s}^2(\boldsymbol{q})+\Delta_0^2(\boldsymbol{q})}}\,\tanh\frac{\sqrt{\varepsilon_{\pm s}^2(\boldsymbol{q})+\Delta_0^2(\boldsymbol{q})}}{2k_B T}.
\end{align}
The sum in moir\'e reciprocal lattice vectors $\mathbf{G}$ accounts for umklapp scattering in the superlattice and the index $\eta$ represents \textit{intra-node} ($\eta=0$) and \textit{inter-node} ($\eta=\pm1$) momentum exchange processes with amplitudes given by\begin{subequations}
\begin{align}
& f_s^{(0)}\left(\boldsymbol{q},\boldsymbol{p},\mathbf{G}\right)=\frac{1}{2}\sum_{\{\boldsymbol{\kappa}_{\nu}\}}\left| \int d\mathbf{r}\, e^{i\left(\boldsymbol{q}-\boldsymbol{p}+\mathbf{G}\right)\cdot\mathbf{r}} \left[u_{\pm s,\boldsymbol{\kappa}_{\nu}+\boldsymbol{q}}\left(\mathbf{r}\right)\right]^*u_{\pm s,\boldsymbol{\kappa}_{\nu}+\boldsymbol{p}}\left(\mathbf{r}\right) \right|^2,\\
&  f_s^{(+1)}\left(\boldsymbol{q},\boldsymbol{p},\mathbf{G}\right)=\frac{1}{2}\left| \int d\mathbf{r}\, e^{i\left(\boldsymbol{q}-\boldsymbol{p}+\mathbf{k}_0+\mathbf{G}\right)\cdot\mathbf{r}}\left[u_{\pm s,\boldsymbol{\kappa}_{b}+\boldsymbol{q}}\left(\mathbf{r}\right)\right]^*u_{\mp s,\boldsymbol{\kappa}_{t}+\boldsymbol{p}}\left(\mathbf{r}\right) \right|^2,\\
& f_s^{(-1)}\left(\boldsymbol{q},\boldsymbol{p},\mathbf{G}\right)=\frac{1}{2}\left| \int d\mathbf{r}\, e^{i\left(\boldsymbol{q}-\boldsymbol{p}-\mathbf{k}_0+\mathbf{G}\right)\cdot\mathbf{r}} \left[u_{\pm s,\boldsymbol{\kappa}_{t}+\boldsymbol{q}}\left(\mathbf{r}\right)\right]^*u_{\mp s,\boldsymbol{\kappa}_{b}+\boldsymbol{p}}\left(\mathbf{r}\right) \right|^2.
\end{align}
\end{subequations}
\end{widetext}

Equation~\eqref{eq:gap_equation} admits a simple solution if only intra-node scattering processes with small momentum exchange are retained.
For a Coulomb potential screened by a double gate, $V\left(\boldsymbol{p}\right)=\frac{e^2}{4\pi\epsilon|\boldsymbol{p}|}\tanh\left(d|\boldsymbol{p}|\right)\approx\frac{e^2 d}{4\pi \epsilon}\equiv V$, this is justified if the separation between gates $d$ is larger than the moir\'e pitch, for in that case $V(\boldsymbol{p}+\mathbf{G})\sim V\times L_{\textrm{m}}/d<V$. In the Dirac approximation, we have $f_{s}^{(0)}\left(\boldsymbol{q},\boldsymbol{p},\mathbf{0}\right)=|\boldsymbol{\tilde{u}}_{\pm s}^{*}\left(\boldsymbol{q}\right)\cdot\boldsymbol{\tilde{u}}_{\pm s}\left(\boldsymbol{p}\right)|^2\approx\frac{1+\cos\left(\theta_{\boldsymbol{q}}-\theta_{\boldsymbol{p}}\right)}{2}$. The order parameter becomes momentum independent. The system is analogous to a $s$-wave superconductor, where the scale determined by the Debye frequency is now substituted by the energy shift proportional to heterostrain. The gap in the limit of zero temperature is
\begin{align}
\label{eq:gap_estimate}
\Delta_0\left(T=0\right)\approx\frac{2g_{A_2}\bar{u}}{\sinh\left(\frac{2}{\nu_FV}\right)},
\end{align}
where $\nu_F$ is the density of states per valley and spin at the Fermi level, $\nu_F\approx\frac{g_{A_2}\bar{u}}{\pi (\hbar v_F^*)^2}$. The mean-field transition temperature is \begin{align}
T_c\approx \frac{2e^{\gamma_E}g_{A_2}\bar{u}}{\pi k_B}e^{-\frac{2}{\nu_FV}},
\end{align}
where $\gamma_E$ is the Euler constant and the extra-factor 2 in the exponent comes from the suppression of backscattering due to the chirality of quasiparticles. Deviations from the neutrality point reduces the nesting between electron and hole Fermi contours, thus reducing the condensation energy. The excitonic insulator disappears at $\mu_c\approx\pm \Delta_0/\sqrt{2}$. By related arguments \cite{SM}, impurity scattering and spatial inhomogeneities of the strain fields have a pair-breaking effect akin to magnetic disorder in superconductors. This sensitivity to disorder might be the cause of the disparity in experimental results.

The excitonic instability may also induce spin correlations in the ground state. In the absence of atomic-scale interactions, the Hamiltonian is invariant under independent spin rotations on each valley sector. The order-parameter manifold is U$_{v}$(2)$\simeq$U$_{v}$(1)$\times$SU$_{v}$(2), with SU$_{v}$(2): $e^{i\frac{\theta_s}{2}\boldsymbol{n}_s\cdot\,\hat{\Lambda}_z\hat{\boldsymbol{s}}}$. In a triplet state ($\theta_s\neq0$), circulating currents polarized along the quantization axis $\boldsymbol{n}_s$ give rise to out-of-phase density waves for opposite spins, which cancel exactly in a \textit{spin flux} phase at $\theta_s=\pi/2$ in this parametrization. The spin sector is thermally disordered in the continuum theory \cite{KT}, while long-range intervalley correlations are limited by the proliferation of vortices. Short-scale interactions will pin the phases $\theta_{v,s}$ and vortex excitations will remain localized.

These are interesting objects, nonetheless. Vortices $e^{i\theta_v}=(x\pm iy)/|x+iy|$ carry fermion numbers, which can be understood in analogy with Kekul\'e bond order in the lowest Landau level of graphene \cite{Kekule1,Kekule2}. Compared to that case, the number of topologically stable \cite{Jackiw-Rossi,Teo-Kane} zero modes localized within the vortex core is duplicated by the symmetry operation formed by time reversal and valley $\pi$-rotations \cite{SM}. While in the Kekul\'e bond order vortices carry anomalous quantum numbers due to charge fractionalization \cite{Kekule1}, in the excitonic insulator the quantum numbers are those of ordinary electrons due to (orbital) Kramers degeneracy. 

In conclusion, the above arguments suggest that spatial homogeneity and transport gaps at neutrality share the same origin: uniform strain fields that pin the twist angle and favor an excitonic instability of the Fermi surface. The prefactor in Eq.~\eqref{eq:gap_estimate} can be as large as 10 meV around the magic angle for realistic values of heterostrain \cite{Fu}. The excitonic gap is very sensitive to the screening of the Coulomb interaction through the exponential dependence on $2/(\nu_F V)\propto \epsilon/d$. Experimentally, this is a common trend in all the insulating states \cite{screening1,screening2,screening3}, while the superconductors seem to be more resilient. The proposed excitonic state breaks time-reversal symmetry and the approximate conservation of charge on each valley. The associate order parameter is a Kekul\'e-like orbital magnetization density wave, which could be directly observed with STM. In addition, the condensate of electron-hole pairs supports neutral valley supercurrents. This specific form of symmetry breaking should be manifested in magnetotransport. In fact, assuming a weak coupling with the encapsulating boron nitride, the Landau level degeneracy deduced from the mean field Hamiltonian $\hat{\mathcal{H}}_0+\hat{\Delta}$ reproduces the sequence $\pm2,\pm4,\pm8,\pm12...$\cite{SM} observed in the insulating devices of Ref.~\onlinecite{ICFO}. Finally, the proposed mechanism leads to a competition between insulating and superconducting phases depending on wether Coulomb repulsion or attractive interactions (mediated by phonons or other collective modes) dominate.
\acknowledgments
\textbf{Acknowledgments.}-- I would like to thank T. Cea, F. Guinea, and C. Rubio-Verd\'u for valuable discussions. This work has been supported by the NSF MRSEC program Grant No. DMR-1420634.

\clearpage
\onecolumngrid

\appendix

\section{Supplementary Material for \\ \textit{Strain-induced excitonic instability in twisted bilayer graphene}}
The relation between the low-energy action in the main text and the continuum models in the literature is clarified. I also derive zero-mode solutions of vortices in the excitonic order parameter, and the degeneracy of Landau levels in mean field.
\maketitle
\onecolumngrid

\section{Symmetry considerations}

Structures formed by two rigid graphene layers can be generically described by a twist angle $\theta$ and a relative translation $\mathbf{u}$. The shift in the periodicities of the two Bravais lattices define the vectors of a beating pattern,\begin{align}
\mathbf{G}_i=\hat{R}_{-\frac{\theta}{2}}\,\mathbf{g}_i-\hat{R}_{\frac{\theta}{2}}\,\mathbf{g}_i=-2\sin\frac{\theta}{2}\,\mathbf{\hat{z}}\times\mathbf{g}_i,
\end{align}
where $\mathbf{g}_i$ are vectors of the original reciprocal lattice (prior to the twist) and $\hat{R}_{\theta}$ is a SO(2) rotation of angle $\theta$ along the vertical axis. The moir\'e pattern is defined by the dual to $\{\mathbf{G}\}$, spanned by vectors
\begin{align}
\label{eq:moire_vectors}
\mathbf{A}_{i}=-\frac{1}{2\sin\frac{\theta}{2}}\,\mathbf{\hat{z}}\times\mathbf{a}_{i},
\end{align}
where $\mathbf{a}_i$ are primitive vectors of the original Bravais lattice, $\mathbf{a}_{i}\cdot\mathbf{g}_j=2\pi\delta_{ij}$.

%Within the moir\'e supercell, the system explores all possible stacking configurations. Let me introduce a function $\mathbf{d}\left(\mathbf{r}\right)$ that relates a given lateral position $\mathbf{r}$ with the local stacking configuration parametrized in terms of the relative translation $\mathbf{d}$ that would generate the same configuration starting from AA stacking. Consider then a point $\mathbf{r}$ and its projections in the top and bottom layer before the shift; I define $\mathbf{d}(\mathbf{r})$ as the difference in lateral positions of the projected points after the shift,
%\begin{align}
%\label{eq:substitution}
%\mathbf{d}:\mathbf{r}\longrightarrow \mathbf{d}\left(\mathbf{r}\right)=2\sin\frac{\theta}{2}\,\mathbf{\hat{z}}\times\mathbf{r}+\mathbf{u}.
%\end{align}
%The local stacking order is defined up to translations on the original Bravais lattice; high-symmetry stackings, AA, AB, and BA, correspond to $\mathbf{d}=\mathbf{0},\boldsymbol{\delta}_i,2\boldsymbol{\delta}_i$, respectively, where $\boldsymbol{\delta}_{1,2,3}$ represent the three vectors connecting nearest neighbors in the honeycomb lattice. It is worth emphasizing that $\mathbf{u}$ determines the global position of high-symmetry stacking configurations with respect to a laboratory coordinate frame, but not their relative positions within the moir\'e supercell. Therefore, 

The relative displacement $\mathbf{u}$ does not affect the commensuration relation between the moir\'e pattern and the atomic graphene lattices, although it does modify the spatial symmetry of the structure.
%In order to verify the fitness of the previous definition, consider the positions of the beating pattern maxima $\mathbf{R}_{\textrm{max}}$ (maximum overlap of thw two layers), which coincide with regions of AA stacking and thus should verify\begin{align}
%\mathbf{d}\left(\mathbf{R}_{\textrm{max}}\right)=n\,\mathbf{a}_1+m\,\mathbf{a}_2,
%\end{align}
%with $n,m$ integers. It follows then from the previous definition that\begin{align}
%\mathbf{R}_{\textrm{max}}=n\,\mathbf{A}_1+m\,\mathbf{A}_2+\mathbf{\tilde{u}},
%\end{align}
%where $\mathbf{\tilde{u}}$ represents the translation of the beating pattern as a result of a relative translation of the layers,\begin{align}
%\label{eq:beating_center}
%\mathbf{\tilde{u}}=\frac{1}{2\sin\frac{\theta}{2}}\,\mathbf{\hat{z}}\times\mathbf{u}.
%\end{align}
%Both in Eqs.~\eqref{eq:moire_vectors}~and~\eqref{eq:beating_center} we have the angle-dependent amplification factor that relates the original lattice period with the moir\'e pitch, $L_{\textrm{m}}$.
As the twist angle decreases, however, the differences between commensurate and incommensurate structures become negligible and the spectrum remains approximately invariant under relative translations of one layer with respect to the other. The point group symmetry is effectively $D_6$, formed by a six-fold rotation axis along a common hexagon center and six in-pane C$_2$ axes that exchange the layers. The moir\'e beating pattern defines an emergent translational symmetry. These are, in fact, the symmetries of the continuum model, which follows from a $k\cdot p$ expansion around the two microscopic valleys (labelled by $\zeta=\pm 1$) of the two layers (labelled by $\nu=t,b$). The other internal number is the sublattice projection of the wave function (labelled by $\alpha=A,B$).

\subsection{Bloch wave functions}

The moir\'e translational symmetry implies that the single-particle Hamiltonian can be diagonalized in a basis of Bloch wave functions. In the continuum model, these are of the form
 \begin{align}
 \label{eq:Bloch}
u_{\lambda,\zeta,\mathbf{q}}\left(\mathbf{r}\right)=\frac{e^{i\mathbf{q}\cdot\mathbf{r}}}{\sqrt{A}}\sum_{\alpha,\nu}\sum_{\left\{\mathbf{G}\right\}}u_{\lambda,\zeta,\mathbf{G}}^{\alpha,\nu}\left(\mathbf{q}\right)f_{\zeta}^{\alpha,\nu}\left(\mathbf{r}\right)e^{i\left(\mathbf{G}-\zeta\boldsymbol{\kappa}_{\nu}\right)\cdot\mathbf{r}},
\end{align}
where $A$ is the area of the system, $\lambda$ is the band index, and $\mathbf{q}$ is the quasi-momentum restricted to the first moir\'e Brillouin zone. Equation~\eqref{eq:Bloch} consists of a superposition of plane waves separated in momentum by vectors of the beating pattern $\mathbf{G}$; the coefficients of this expansion describe the modulation of envelope wave functions within the moir\'e cell. The factors $f_{\zeta}^{\alpha,\nu}(\mathbf{r})$ describe the fast (on the atomic scale) modulation of the wave function. In a tight-binding description, which is usually the starting point for the derivation of the model, these are given by
\begin{align}
\label{eq:f_factor}
f_{\zeta}^{\alpha,\nu}(\mathbf{r})=\frac{1}{\sqrt{N}}\sum_{i}e^{i\mathbf{K}_{\zeta}^{(\nu)}\cdot\mathbf{R}_{i}^{\alpha,\nu}}\,\Phi^{\alpha,\nu}\left(\mathbf{r}-\mathbf{R}_{i}^{\alpha,\nu}\right),
\end{align}
where $\mathbf{R}_{i}^{\alpha,\nu}$ represent the positions of the atoms, the sum is extended to $N$ microscopic cells, and $\Phi^{\alpha,\nu}(\mathbf{r})$ are Wannier functions of $\pi$ orbitals in $\alpha$ sublattice of layer $\nu$; $\mathbf{K}_{\zeta}^{(\nu)}$ represents the positions of the microscopic valleys at opposite corners of the graphene Brillouin zone. Due to the relative twist, the valleys of the two layers are shifted by a vector $\pm\mathbf{k}_0\equiv\mathbf{K}_{\pm}^{(b)}- \mathbf{K}_{\pm}^{(t)}$, where $|\mathbf{k}_0|=2|\mathbf{K}_{\pm}^{(\nu)}|\sin\frac{\theta}{2}$; this shift, along with the coupling between layers, defines the moir\'e superlattice potential.

In the presence of moir\'e translational symmetry, microscopic valleys are folded back to the corners of the moir\'e Brillouin zone, $\mathbf{K}_{\zeta}^{(\nu)}\equiv\zeta\boldsymbol{\kappa}_{\nu}$. This folding scheme is only exact for a subset of commensurate angles, although the continuum model neglects these details on the atomic scale. By imposing this scheme we enforce the approximate translational symmetry, $u_{\lambda,\zeta,\mathbf{q}}\left(\mathbf{r}+\mathbf{R}\right)=e^{i\mathbf{q}\cdot\mathbf{R}}\,u_{\lambda,\zeta,\mathbf{q}}\left(\mathbf{r}\right)$, with $\mathbf{R}$ spanned by $\mathbf{A}_{1,2}$. Note that this symmetry allows for interlayer Bragg scattering between opposite valleys, but these processes are negligible for small twist angles as the pitch of the moir\'e pattern is large compared to the carbon-carbon distance. For this reason, eigenstates~\eqref{eq:Bloch} can labelled by the valley index $\zeta$, expressing the separate conservation of charge within each valley. We end up then with two valley sectors connected by time-reversal symmetry, $\mathcal{T}$.

As some of the operations in the point group exchange the valleys, we may consider instead new anti-unitary symmetries formed by those in combination with time-reversal symmetry. These new operations along with the subgroup $D_3$ generated by C$_{3z}$ and C$_{2x}$ rotations form the magnetic group $D_6(D_3)$ operating within a single valley sector. The group of the wave vectors $\mathbf{q}=\boldsymbol{\kappa}_{\nu}$ is $C_6(C_3)$, whose generators are C$_{3z}$ and C$_{2z}\mathcal{T}$. The bands at these points can be classified according to the irreducible representations of the unitary subgroup $C_3$, which are all one dimensional with characters of the form $e^{\frac{i2\pi\lambda}{3}}$, with $\lambda=0,1,2$. For this magnetic group, it can be shown that eigenstates belonging to complex representations $\lambda=1,2$ are Kramers degenerate, i.e., they are connected by the anti-unitary operation C$_{2z}\mathcal{T}$. This is the case of the lowest-energy eigenstates dominated by the original Dirac points, where the non-trivial transformation under C$_{3z}$ rotations originates from the fast-oscillating factors $f_{\zeta}^{\alpha,\nu}\left(\mathbf{r}\right)$. As a result, the Dirac crossings folded onto $\boldsymbol{\kappa}_{\nu}$ are preserved, their chirality being determined by the valley index $\zeta$. In the presence of C$_{2x}$ symmetry (broken by heterostrain or layer-asymmetric perturbations) the Dirac points at the two inequivalent $\boldsymbol{\kappa}_{\nu}$ points must be degenerate in energy. The extra valley degeneracy follows from $\mathcal{T}$ (not contained in the magnetic group). 

\subsection{Representations and matrix algebra}

Any operator in the Hilbert space associated with these four Dirac points can be expanded in a basis of 64 independent $8\times8$ matrices, the identity $\hat{1}$ and all the possible combinations of elements in three inter-commuting Pauli algebras introduced in the main text. The algebraical relations between these operators can be determined from representation theory without relaying on a specific basis.

Let us define $\hat{\Sigma}_z$ as the generator of C$_{3z}$ rotations,\begin{align}
\label{eq:C3}
\text{C$_{3z}$}:\,\, e^{i\frac{2\pi}{3}\hat{\Sigma}_z}.
\end{align}
$\Sigma_z$ must transform as a z-component of angular momentum, $\hat{\Sigma}_z\sim A_2$. The other two matrices are chosen to form a doublet $(\hat{\Sigma}_x,\hat{\Sigma}_y)\sim E_1$. The Pauli matrix algebra follows from the reduction of matrix products into irreducible representations as $E_1\times E_1\sim A_1+A_2+E_2$.

The operator $\hat{\Lambda}_z$ is defined as the generator of U$_{v}$(1) rotations (separate charge conservation on each valley),\begin{align}
\label{eq:U_valley}
\text{U$_{v}$(1)}:\,\, e^{\frac{i\theta_v}{2}\hat{\Lambda}_z}.
\end{align}
It must belong to a $B_1$ representation. Then, Pauli matrix algebra and invariance under C$_{3z}$ rotations imply $\hat{\Lambda}_x\sim A_2$, $\hat{\Lambda}_y\sim B_2$. This choice corresponds to a gauge in which C$_{2y}$ is represented by a real symmetric matrix,\begin{align}
\text{C}_{2y}:\,\,\, \hat{\Sigma}_y\hat{\Lambda}_y.
\end{align}

\begin{table}
\centering
\begin{tabular}{|c||c|c|c|c|c|c|c|c|c|}
\hline
$D_6''$ & $E$ & 2 $T$ & 3 $T\times$C$_{2z}$ & 2 C$_{3z}$ & 4 $T\times$C$_{3z}$ & 6 $T\times$C$_{6z}$ & 9 $T\times$C$_{2x}$ & 3 C$_{2y}$ & 6 $T\times$C$_{2y}$ \\
\hline
\hline
$A_1$ & 1 & 1 & 1 & 1 & 1 & 1 & 1 & 1 & 1 \\
\hline
$A_2$ & 1 & 1 & 1 & 1 & 1 & 1 & -1 & -1 & -1  \\
\hline
$B_1$ & 1 & 1 & -1 & 1 & 1 & -1 & 1 & -1 & -1  \\
\hline
$B_2$ & 1 & 1 & -1 & 1 & 1 & -1 & -1 & 1 & 1  \\
\hline
$E_1$ & 2 & 2 & -2 & -1 & -1 & 1 & 0 & 0 & 0  \\
\hline
$E_2$ & 2 & 2 & 2 & -1 & -1 & -1 & 0 & 0 & 0  \\
\hline
$E_1'$ & 2 & -1 & 0 & 2 & -1 & 0 & 0 & 2 & -1  \\
\hline
$E_2'$ & 2 & -1 & 0 & 2 & -1 & 0 & 0 & -2 & 1  \\
\hline
$G$ & 4 & -1 & 0 & -2 & 1 & 0 & 0 & 0 & 0  \\
\hline
\end{tabular}
\caption{Character table of $D_6''=D_6+T_{\mathbf{A}_1}\times D_6+T_{\mathbf{A}_2}\times D_6$. The numbers in the first row indicates the number of operations within a given class (36 in total, 12 of the original point group and the new operations resulting from a point group transformation followed by an elementary translation).}
\label{tab:character_table}
\end{table}

The operator $\hat{\Gamma}_z$ is the generator of moir\'e translations; elementary translations of displacement $\mathbf{A}_{1,2}$ acting on the Bloch wave functions with $\mathbf{q}=\boldsymbol{\kappa}_{\nu}$ form a cyclic subgroup of order 3 given by matrices\begin{align}
\label{eq:translations}
T_{\mathbf{A}_{1,2}}:\,\,\, e^{\pm\frac{i2\pi}{3}\hat{\Gamma}_z}.
\end{align}
$\hat{\Gamma}_z$ belongs to the remaining one-dimensional representation, $B_2$.

With this information, we can already reproduce the table in the main text. For the remaining 32 operators, instead of dealing with $D_6$, it is more convenient to factorize out the elementary translations $T_{\mathbf{A}_{1,2}}$ from the moir\'e translation group and integrate them into the point group, which becomes $D_6''=D_6+T_{\mathbf{A}_1}\times D_6+T_{\mathbf{A}_2}\times D_6$. On more physical grounds, what we are doing is to consider translations in the tripled supercell spanned by primitive vectors $\mathbf{A}_1+\mathbf{A}_2$ and $2\mathbf{A}_2-\mathbf{A}_1$, so inequivalent corners $\boldsymbol{\kappa}_{\nu}$ are folded into the $\boldsymbol{\gamma}$ point. The irreducible representations and character table of $D_6''$ can be easily constructed just by noting that operations of $D_6$ in the little group of the wave vectors $\boldsymbol{\kappa}_{\nu}$, i.e., the subgroup $D_3$ generated by C$_{3z}$ and C$_{2y}$ rotations, are in different conjugacy classes that the same operations followed by an elementary translation. This implies that there are three new classes in $D_6''$, and therefore three new irreducible representations denoted by $E_1'$, $E_2'$ and $G$;  $E_{1,2}'$ are 2-dimensional and $G$ is 4-dimensional, which follows from the fact that $2^2+2^2+4^2=24$ must equal the number of new operations consisting of any of the twelve operations in $D_6$ followed by an elementary translation. Similar algebraical relations fix the characters of the representations, displayed in Table~\ref{tab:character_table}. Operators $\hat{\Gamma}_{x,y}$ belong to one of the new doublets; since $\boldsymbol{\kappa}_{\nu}$ points remain invariant under C$_{2y}$, it can only be $(\hat{\Gamma}_x,\hat{\Gamma}_y)\sim E_1'$.
This also fixes the representations of the remaining two-fold rotations,\begin{subequations}
\label{eq:C2}
\begin{align}
& \text{C}_{2x}:\,\,\, \hat{\Sigma}_x\hat{\Lambda}_z\hat{\Gamma}_x,\\
&  \text{C}_{2z}:\,\,\, \hat{\Sigma}_z\hat{\Lambda}_x\hat{\Gamma}_x.
\end{align}
\end{subequations}
The classification of the rest of operators are shown in Table~\ref{tab:operators2}.

Equation~(1) of the main text forms a basis for the representations of $D_6''$ introduced above. Under an operation $g\in D_{6}''$, the 8-component fermion operators of the low-energy theory are transformed as
\begin{align}
\boldsymbol{\hat{\psi}}\left(\tau,\mathbf{r}\right)\overset{g}{\longrightarrow} \hat{\mathcal{U}}_{g}\cdot\boldsymbol{\hat{\psi}}\left(\tau,\hat{R}_{g}^{-1}\mathbf{r}\right),
\end{align}
where $\hat{R}_{g}$ is the representation in coordinate space and $\hat{\mathcal{U}}_{g}$ are unitary matrices given by Eqs.~\eqref{eq:C3}-\eqref{eq:translations}-\eqref{eq:C2} and their products following the multiplication table of the group. The matrix basis is given by %. More specifically, if Dirac spinors of opposite chirality $\zeta=\pm 1$ are defined as
%\begin{subequations}
%\begin{align}
%\vec{u}_{\zeta,\boldsymbol{\kappa}_{\nu}}\left(\mathbf{r}\right)=\left[u_{1,\zeta,\boldsymbol{\kappa}_{\nu}}\left(\mathbf{r}\right),\zeta\, u_{2,\zeta,\boldsymbol{\kappa}_{\nu}}\left(\mathbf{r}\right)\right]^T,
%\end{align}
 %\end{subequations}
%where $u_{\lambda,\zeta,\boldsymbol{\kappa}_{\nu}}\left(\mathbf{r}\right)$ are Bloch wave functions~\eqref{eq:Bloch} with band index $\lambda$ labelling the complex irreducible representation of $C_3$, then we have
\begin{subequations}\begin{align}
& \hat{\Sigma}_i=\hat{1}\otimes\hat{1}\otimes\hat{\lambda}_i,\\
& \hat{\Lambda}_i=\hat{1}\otimes\hat{\zeta}_i\otimes\hat{1},\\
& \hat{\Gamma}_i=\hat{\kappa}_i\otimes\hat{1}\otimes\hat{1},
\end{align}
\end{subequations}
where $\hat{\lambda}_i$, $\hat{\zeta}_i$, and $\hat{\kappa}_i$ are Pauli matrices acting on the corresponding degrees of freedom (isospin, valley and mini-valley). In this gauge, time reversal symmetry is implemented by the anti-unitary operator
\begin{align}
\mathcal{T}:\,\,\hat{\Sigma}_y\hat{\Lambda}_y\hat{\Gamma}_x\mathcal{K},
\end{align}
where $\mathcal{K}$ is complex conjugation. %U$_v$(1) matrices are given by Eq.~\eqref{eq:U_valley}.

\begin{table}
\centering
\begin{tabular}{|c||c|c|c|}
\hline
valley & $E_1'\sim\left[A_1,B_2\right]$ & $E_2'\sim\left[A_2,B_1\right]$ & $G\sim\left[E_1,E_2\right]$  \\
\hline
\hline
$0$ & $\left[\begin{array}{c}\hat{\Gamma}_x \\ \hat{\Gamma}_y \end{array}\right]$ (+) & $\left[\begin{array}{c}\hat{\Sigma}_z\hat{\Gamma}_x \\ \hat{\Sigma}_z\hat{\Gamma}_y \end{array}\right]$ (-) & $\left[\hat{\Sigma}_x\hat{\Gamma}_x,\hat{\Sigma}_y\hat{\Gamma}_x,-\hat{\Sigma}_y\hat{\Gamma}_y,\hat{\Sigma}_x\hat{\Gamma}_y\right]$ (-)  \\
\hline
$x$ & $\left[\begin{array}{c}\hat{\Sigma}_z\hat{\Lambda}_x\hat{\Gamma}_x \\ \hat{\Sigma}_z\hat{\Lambda}_x\hat{\Gamma}_y \end{array}\right]$ (+) & $\left[\begin{array}{c}\hat{\Lambda}_x\hat{\Gamma}_x \\ \hat{\Lambda}_x\hat{\Gamma}_y\end{array}\right]$ (-) & $\left[-\hat{\Sigma}_y\hat{\Lambda}_x\hat{\Gamma}_x,\hat{\Sigma}_x\hat{\Lambda}_x\hat{\Gamma}_x,\hat{\Sigma}_x\hat{\Lambda}_x\hat{\Gamma}_y,\hat{\Sigma}_y\hat{\Lambda}_x\hat{\Gamma}_y\right]$ (+)  \\
\hline
$y$ & $\left[\begin{array}{c} -\hat{\Lambda}_y\hat{\Gamma}_y \\ \hat{\Lambda}_y\hat{\Gamma}_x \end{array}\right]$ (-) & $\left[\begin{array}{c}-\hat{\Sigma}_z\hat{\Lambda}_y\hat{\Gamma}_y \\ \hat{\Sigma}_z\hat{\Lambda}_y\hat{\Gamma}_x \end{array}\right]$ (+) & $\left[\hat{\Sigma}_x\hat{\Lambda}_y\hat{\Gamma}_y,\hat{\Sigma}_y\hat{\Lambda}_y\hat{\Gamma}_y,\hat{\Sigma}_y\hat{\Lambda}_y\hat{\Gamma}_x,-\hat{\Sigma}_x\hat{\Lambda}_y\hat{\Gamma}_x\right]$ (+)  \\
\hline
$z$ & $\left[\begin{array}{c}-\hat{\Sigma}_z\hat{\Lambda}_z\hat{\Gamma}_y \\ \hat{\Sigma}_z\hat{\Lambda}_z\hat{\Gamma}_x \end{array}\right]$ (+) & $\left[\begin{array}{c}-\hat{\Lambda}_z\hat{\Gamma}_y \\ \hat{\Lambda}_z\hat{\Gamma}_x \end{array}\right]$ (-) & $\left[\hat{\Sigma}_y\hat{\Lambda}_z\hat{\Gamma}_y,-\hat{\Sigma}_x\hat{\Lambda}_z\hat{\Gamma}_y,\hat{\Sigma}_x\hat{\Lambda}_z\hat{\Gamma}_x,\hat{\Sigma}_y\hat{\Lambda}_z\hat{\Gamma}_x\right]$ (+)  \\
\hline
\end{tabular}
\caption{Hermitian operators mixing the mini-valleys.}
\label{tab:operators2}
\end{table}

\section{Continuum model}

The phenomenological coefficients of the low-energy theory can be estimated from the matrix elements of the Hamiltonian of the continuum model in the Bloch wave basis introduced in Eq.~(1) of the main text. Let me write the total Hamiltonian (now in a larger Hilbert space to be specified next) as $H=H_0+H_{C}$, where $H_0$ includes all the single-particle terms, and $H_C$ represents the long-range Coulomb interaction. In second quantization, we can write the former as\begin{align}
H_0=\sum_{\zeta=\pm 1}\int d\mathbf{r}\, \hat{\psi}_{\zeta}^{\dagger}\left(\mathbf{r}\right)\hat{\mathcal{H}}_{\zeta}\left(\mathbf{r}\right) \hat{\psi}_{\zeta}\left(\mathbf{r}\right),
\end{align}
where $\hat{\psi}_{\zeta}$ are 4-component field operators in sublattice and layer spaces defined in a given valley $\zeta$. Note that the microscopic field operator is given by
\begin{align}
    \hat{\Psi}\left(\mathbf{r}\right)=\sum_{\alpha,\nu}\sum_{\zeta=\pm 1}f_{\zeta}^{\alpha,\nu}\left(\mathbf{r}\right)\,\hat{\psi}_{\zeta}^{\alpha,\nu}\left(\mathbf{r}\right).
\end{align}
$\hat{\mathcal{H}}_{\zeta}\left(\mathbf{r}\right)$ is a smooth varying (on the atomic scale) $4\times 4$ matrix,\begin{align}
    \label{eq:Hamiltonian}
   \hat{\mathcal{H}}_{\zeta}\left(\mathbf{r}\right)=\left[\begin{array}{cc}
  \hat{\mathcal{H}}_{\zeta}^{(t)}\left(\mathbf{r}\right) & \hat{T}_{\zeta}\left(\mathbf{r}\right)\\
  \hat{T}_{\zeta}^{\dagger}\left(\mathbf{r}\right) & \hat{\mathcal{H}}^{(b)}_{\zeta}\left(\mathbf{r}\right)
   \end{array}\right].
\end{align}

In the absence of strain fields, the blocks in the diagonal are Dirac Hamiltonians describing electrons in the top and bottom layers,\begin{align}
\hat{\mathcal{H}}_{\zeta}^{(t/b)}\left(\mathbf{r}\right)=-i\hbar v_F\cos\frac{\theta}{2}\,\hat{\boldsymbol{\sigma}}_{\zeta}\cdot\boldsymbol{\partial}\mp i\hbar v_F\sin\frac{\theta}{2}\left(\hat{\boldsymbol{\sigma}}_{\zeta}\times\boldsymbol{\partial}\right)_z,
\end{align}
where $\boldsymbol{\hat{\sigma}}_{\zeta}=(\zeta \hat{\sigma}_x,\hat{\sigma}_y)$ is a vector of Pauli matrices acting on spin and $\hbar v_F=\sqrt{3} ta/2$, $t$ being the intra-layer hopping parameter. It is worth emphasizing that derivatives act only on the coordinates of the envelope wave function. Hereafter $x$ and $y$ coordinates lie along C$_{2}$ axes; the sublattice basis on each layer is properly adjusted by the twist-dependent factors. The interlayer tunneling terms are \begin{subequations}
\label{eq:tunneling}
\begin{align}
& \hat{T}_{\zeta}\left(\mathbf{r}\right)=w\sum_{n=0,1,2}\hat{T}_{\zeta}^{(n)}\, e^{i\zeta\mathbf{k}_n\cdot\mathbf{r}},
\end{align}
with matrices \begin{align}
\label{eq:T0}
\hat{T}_{\zeta}^{(n)}= e^{i\zeta\frac{n2\pi}{3}\hat{\sigma}_z}\,\hat{T}_{0}\,e^{-i\zeta\frac{n2\pi}{3}\hat{\sigma}_z},\,\,\,\hat{T}_{0}=\hat{1}+\hat{\sigma}_x,
\end{align}
and vectors\begin{align}
& \mathbf{k}_0=\boldsymbol{\kappa}_b-\boldsymbol{\kappa}_t,\\
& \mathbf{k}_1=\boldsymbol{\kappa}_b-\boldsymbol{\kappa}_t+\mathbf{G}_2,\\
& \mathbf{k}_2=\boldsymbol{\kappa}_b-\boldsymbol{\kappa}_t-\mathbf{G}_1.
\end{align}
\end{subequations}
In this approximation, the interlayer tunneling rate $w$ is uniform over the moir\'e cell, and the spatial modulation has a purely geometrical origin associated with momentum boosts due to the relative displacements of the Dirac points on each layer.

Invariance under $g=$ C$_{3z}$, C$_{2x}$ implies\begin{align}
\hat{\mathcal{H}}_{\zeta}\left(\mathbf{r}\right)=\hat{\mathcal{U}}_{g}\,\hat{\mathcal{H}}_{\zeta}\left(\hat{R}^{-1}_g\mathbf{r}\right)\,\hat{\mathcal{U}}_{g}^{\dagger},
\end{align}
where $\hat{\mathcal{U}}_{g}$ is the unitary representation in layer$\otimes$sublattice space:\begin{subequations}\begin{align}
& \text{C$_{3z}$}:\,\, e^{i\zeta\frac{2\pi}{3}\hat{\sigma}_z},\\
& \text{C$_{2x}$}:\hat{\nu}_x\otimes\hat{\sigma}_x.
\end{align}
\end{subequations}
Here I have introduced Pauli matrices $\hat{\nu}_i$ acting on layer indices. The complex representation of C$_{3z}$ rotations is reminiscent of the fast-oscillating factor of the wave functions, $f_{\zeta}^{\alpha,\nu}(\mathbf{r})$.

For operations that exchange the valleys, $g=$ C$_{3z}$, C$_{2x}$,\begin{align}
\hat{\mathcal{H}}_{-\zeta}\left(\mathbf{r}\right)=\hat{\mathcal{U}}_{g}\,\hat{\mathcal{H}}_{\zeta}\left(\hat{R}^{-1}_{g}\mathbf{r}\right)\,\hat{\mathcal{U}}_{g}^{\dagger},
\end{align}
with
\begin{subequations}
\begin{align}
& \text{C$_{2z}$}:\,\, \hat{\sigma}_{x},\\
& \text{C$_{2y}$}:\,\, \hat{\nu}_{x}.
\end{align}
\end{subequations}
The form of these operators and their counterparts in the low-energy theory introduced in the previous section can be reconciled by noting that quantum numbers $\lambda$ are associated with predominant sublattice polarization of the Bloch wave function (opposite for each valley due to the inverted chirality), whereas the combination of valley and mini-valley indices gives an idea of the predominant layer polarization dictated by the folding scheme.

The tunneling matrix $\hat{T}_0$ in Eq.~\eqref{eq:T0} admits a more general parametrization compatible with $D_6$ symmetry,\begin{align}
\label{eq:generalization}
\hat{T}_0=\beta\hat{1}+\hat{\sigma}_x.
\end{align}
In the calculations of Fig.~2 of the main text I took $w=97.5$ meV, $\beta=0.82$. A diminished interlayer hopping in AA stacked regions accounts for lattice relaxation in a phenomenological manner. If $\beta=0$, the index $\lambda$ can be identified directly with sublattice polarization. This is the result of a chiral symmetry different from the one discussed in the main text. The latter results from the approximate electron-hole symmetry of the spectrum at small twist angles. If we neglect the spinor basis rotation in the diagonal blocks, then the Hamiltonian of the continuum model anti-commutes with the following anti-unitary operation:\begin{align}
\label{eq:P_def}
\mathcal{P}:\,\,\, i\,\hat{\nu}_y\otimes\hat{\sigma}_x\,\mathcal{K}.
\end{align}
In the low-energy subspace, the latter reads
\begin{align}
\mathcal{P}:\,\, i\,\hat{\Sigma}_x\hat{\Gamma}_y\mathcal{K},
\end{align}
which, combined with $\mathcal{T}$, defines the chiral operator in Eq.~(2) of the main text.

\subsection{Plane waves}

The Hamiltonian can be diagonalized in a basis of operators of the form\begin{align}
    \hat{c}_{\zeta}^{\alpha,\nu}\left(\mathbf{k}\right)=\frac{1}{\sqrt{A}}\int d\mathbf{r}\, e^{-i\left(\mathbf{k}-i\zeta\boldsymbol{\kappa}_{\nu}\right)\cdot\mathbf{r}}\,\hat{\psi}_{\zeta}^{\alpha,\nu}\left(\mathbf{r}\right).
\end{align}
These operators annihilate plane waves of momentum $\mathbf{k}$ referred to the folded position of the corresponding valley $\zeta$ but not restricted to the first Brillouin zone of the moir\'e superlattice. Nevertheless, we can always write $\mathbf{k}=\mathbf{q}+\mathbf{G}$, with $\mathbf{q}$ within the moir\'e Brillouin zone, and enforce such distinction in the Hilbert space by writing\begin{align}
\label{eq:plane_waves}
\left[\hat{c}_{\zeta}^{\alpha,\nu}\left(\mathbf{q}+\mathbf{G}\right)\right]^{\dagger}\left|0\right\rangle\equiv \left|\mathbf{q}\right\rangle\otimes \left|\mathbf{G},\zeta,\alpha,\nu\right\rangle,
\end{align}
such that
\begin{subequations}\begin{align}
& \left\langle \mathbf{r}\right|\mathbf{q}\rangle=\frac{e^{i\mathbf{q}\cdot\mathbf{r}}}{\sqrt{N_{\textrm{m}}}},\\
& \left\langle \mathbf{r}\right|\mathbf{G},\zeta,\alpha,\nu\rangle=\frac{e^{i\left(\mathbf{G}-\zeta\boldsymbol{\kappa}_{\nu}\right)\cdot\mathbf{r}}}{\sqrt{A_{\textrm{m}}}},
\end{align}\end{subequations}
where $N_{\textrm{m}}$ is the number of moir\'e supercells and $A_{\textrm{m}}$ is their area. The microscopic expression of a plane wave with momentum $\mathbf{k}=\mathbf{q}+\mathbf{G}$ around valley $\zeta$ projected on sublattice $\alpha$ of layer $\nu$ is then\begin{align}
\label{eq:plane_waves_wf}
\Psi_{\zeta,\mathbf{q}+\mathbf{G}}^{\alpha,\nu}\left(\mathbf{r}\right)=f_{\zeta}^{\alpha,\nu}\left(\mathbf{r}\right)\times\left\langle \mathbf{r}\right|\mathbf{q}\rangle\times\left\langle \mathbf{r}\right|\mathbf{G},\zeta,\alpha,\nu\rangle.
\end{align} 
Each of these factors control the spatial dependence of the wave functions on the different length scales of the problem.

As mentioned in the main text, Hartree-Fock calculations predict a variety of possible ground states with small energy differences between them. This is a consequence of the the multiscale nature of the problem. Heterostrain (soon to be introduced in the continuum model) freezes some electronic degrees of freedom and introduces a finite density of states for others, thus defining a natural length/energy scale separation while preserving the electron-hole symmetry of the spectrum. This ultimately justifies the use of an effective action in the main text. The form of the excitonic order parameter, however, is not totally inmune to variations of charge on length scales smaller than the moir\'e pitch; on the contrary, it is connected to these through the definition of the chiral operator in Eq.~(2) of the main text, which comes from the projection of a symmetry in the enlarged Hilbert space of the continuum model. In other words, the correlations between different orbital degrees of freedom (sublattice, layer, valley) are not the result of an emergent symmetry of the low-energy bands around $\boldsymbol{\kappa}_{\nu}$ points; in fact, the chiral symmetry dictating the form of these correlations is a property of the continuum Hamiltonian governing the electronic spectrum at larger energy scales. I will come later to this point with a more specific example regarding the role of \textit{intra-} and \textit{inter-node} Coulomb scattering in defining the order parameter.

The Bloch wave functions~\eqref{eq:Bloch} are just suitable linear combinations of the plane waves in Eq.~\eqref{eq:plane_waves_wf}, where the harmonics on the moir\'e reciprocal lattice are treated as a new internal quantum number; in ket notation,\begin{align}
\label{eq:ket_ansatz}
\left|u_{\lambda,\zeta}\left(\mathbf{q}\right) \right\rangle= \sum_{\alpha}\sum_{\nu}\sum_{\left\{ \mathbf{G}\right\}} u_{\lambda,\zeta,\mathbf{G}}^{\alpha,\nu}\left(\mathbf{q}\right)\left|\mathbf{G},\zeta,\alpha,\nu\right\rangle.
\end{align}
The Hamiltonian can be written as\begin{align}
H_0=\sum_{\zeta=\pm 1}\sum_{\mathbf{q}\in\textrm{mBZ}}\hat{\mathcal{H}}_{\zeta}\left(\mathbf{q}\right)\otimes \left|\mathbf{q}\right\rangle\left \langle \mathbf{q}\right|,
\end{align}
where \begin{align}
\label{eq:matrix_Hamiltonian}
\hat{\mathcal{H}}_{\zeta}\left(\mathbf{q}\right)=&\sum_{\alpha,\beta}\sum_{\left\{\mathbf{G}\right\}} \sum_{\nu}\hbar v_F\left(\mathbf{q}-\zeta\boldsymbol{\kappa}_{\nu}+\mathbf{G}\right)\cdot \left[\hat{\boldsymbol{\sigma}}_{\zeta}^{(\nu)}\right]_{\alpha\beta} \left|\mathbf{G},\zeta,\alpha,\nu\right\rangle\left\langle\mathbf{G}, \zeta,\beta,\nu \right|\\
& +w\left\{\sum_{\alpha,\beta}\sum_{\left\{\mathbf{G}\right\}}\sum_{n} \left[\hat{T}_{\zeta}^{(n)}\right]_{\alpha\beta}\left|\mathbf{G}+\zeta\mathbf{Q}_n,\zeta,\alpha,t\right\rangle\left\langle \mathbf{G}, \zeta,\beta,b \right| + \textrm{h.c.}\right\},
\nonumber
\end{align}
with $\mathbf{Q}_0=\mathbf{0}$, $\mathbf{Q}_1=\mathbf{G}_2$, $\mathbf{Q}_2=-\mathbf{G}_1$ and\begin{align}
\hat{\boldsymbol{\sigma}}_{\zeta}^{(t/b)}=\cos\frac{\theta}{2}\,\hat{\boldsymbol{\sigma}}_{\zeta}\pm \sin\frac{\theta}{2}\boldsymbol{\hat{z}}\times\hat{\boldsymbol{\sigma}}_{\zeta}.
\end{align}
The bands $\varepsilon_{\lambda,\zeta}(\mathbf{q})$ and Bloch wave functions follow from the diagonalization of this matrix,\begin{align}
\hat{\mathcal{H}}_{\zeta}\left(\mathbf{q}\right)\left|u_{\lambda,\zeta}\left(\mathbf{q}\right) \right\rangle=\varepsilon_{\lambda,\zeta}(\mathbf{q})\left|u_{\lambda,\zeta}\left(\mathbf{q}\right) \right\rangle.
\end{align}
The second quantization operator creating an electron with quasi-momentum $\mathbf{q}$ in band $\lambda$ and valley $\zeta$ is defined as\begin{align}
\hat{c}_{\lambda,\zeta}^{\dagger}\left(\mathbf{q}\right)\left|0\right\rangle=\left|\mathbf{q}\right\rangle\otimes\left| u_{\lambda,\zeta}\left(\mathbf{q}\right) \right\rangle,
\end{align}
and therefore,\begin{subequations}
\label{eq:changing_basis_2nd}
\begin{align}
& \hat{c}_{\lambda,\zeta}^{\dagger}\left(\mathbf{q}\right)= \sum_{\alpha}\sum_{\nu}\sum_{\left\{ \mathbf{G}\right\}} u_{\lambda,\zeta,\mathbf{G}}^{\alpha,\nu}\left(\mathbf{q}\right)\,\left[\hat{c}_{\zeta}^{\alpha,\nu}\left(\mathbf{q}+\mathbf{G}\right)\right]^{\dagger},\\
& \hat{c}_{\zeta}^{\alpha,\nu}\left(\mathbf{q}+\mathbf{G}\right)=\sum_{\lambda} u_{\lambda,\zeta,\mathbf{G}}^{\alpha,\nu}\left(\mathbf{q}\right)\hat{c}_{\lambda,\zeta}\left(\mathbf{q}\right).
\end{align}
\end{subequations}
These relations between operators are univocal as long as we define boundary conditions on reciprocal space. If an infinite number of Fourier harmonics is included in the Hamiltonian, then the associated matrix in reciprocal space satisfies\begin{align}
\label{eq:translated_Hamiltonian}
\hat{\mathcal{H}}_{\zeta}\left(\mathbf{q}+\mathbf{G}_i\right)=\hat{\mathcal{U}}_{\mathbf{G}_i}\hat{\mathcal{H}}_{\zeta}\left(\mathbf{q}\right)\hat{\mathcal{U}}_{\mathbf{G}_i}^{\dagger},
\end{align}
where\begin{align}
\hat{\mathcal{U}}_{\mathbf{G}_i}=\sum_{\alpha}\sum_{\nu}\sum_{\left\{ \mathbf{G}\right\}} \left|\mathbf{G}-\mathbf{G}_i,\zeta,\alpha,\nu\right\rangle\left\langle \mathbf{G}, \zeta,\alpha,\nu \right|.
\end{align}
If $|u_{n,\zeta}(\mathbf{q}) \rangle$ is an eigenvector of $\hat{\mathcal{H}}_{\zeta}(\mathbf{q})$ with eigenvalue $\varepsilon_{n,\zeta}\left(\mathbf{q}\right)$, then $\hat{\mathcal{U}}_{\mathbf{G}_i}|u_{n,\zeta}(\mathbf{q}) \rangle$ is an eigenvector of $\hat{\mathcal{H}}_{\zeta}(\mathbf{q}+\mathbf{G}_i)$ with the same eigenvalue. Periodic boundary conditions amounts to the identification\begin{align}
\left|u_{\lambda,\zeta}\left(\mathbf{q}+\mathbf{G}_i\right) \right\rangle\equiv \hat{\mathcal{U}}_{\mathbf{G}_i} \left|u_{\lambda,\zeta}\left(\mathbf{q}\right) \right\rangle \Longrightarrow u_{\lambda,\zeta,\mathbf{G}}^{\alpha,\nu}\left(\mathbf{q}+\mathbf{G}_i\right)=u_{\lambda,\zeta,\mathbf{G}+\mathbf{G}_i}^{\alpha,\nu}\left(\mathbf{q}\right),
\end{align}
or in second quantization,\begin{align}
\hat{c}_{\lambda,\zeta}\left(\mathbf{q}+\mathbf{G}_i\right)=\hat{c}_{\lambda,\zeta}\left(\mathbf{q}\right).
\end{align}

\subsection{Heterostrain}

Consider the strain tensor on each layer, defined as the symmetrized derivative of the corresponding (in-plane) displacement field with respect to the equilibrium position of carbon atoms in the absence of interlayer couplings,\begin{align}
u_{ij}^{\nu}=\frac{1}{2}\left(\partial_i u_j^{\nu}+\partial_j u_i^{\nu}\right).
\end{align}
As in-plane C$_{2}$ symmetries exchange the layers, it is more sensible to consider symmetric and anti-symmetric combinations of the displacements of the two layers. \textit{Heterostrain} refers to the latter. The components of the heterostrain tensor can be arranged in irreducible representations of $D_6$ as\begin{subequations}\begin{align}
& \omega_{xx}+\omega_{yy}\sim A_2,\\
& \left[\begin{array}{cc}
2\omega_{xy}\\
\omega_{xx}-\omega_{yy}
\end{array}
\right]\sim E_2.
\end{align}
\end{subequations}
These, along with time reversal and U$_{\zeta}$(1) valley symmetries dictate the form of the couplings in Eq.~(4) of the main text.

From the definition of the heterostrain tensor, it is implicit that we are not including the effect of lattice relaxation at this level. In fact, we are only interested in the long-scale heterostrain fields that give rise to distortions of the moir\'e pattern. These can be described by changes in the moir\'e reciprocal lattice vectors,\begin{align}
\mathbf{G}_i'\approx-2\sin\frac{\theta}{2}\,\mathbf{\hat{z}}\times\mathbf{g}_i+\hat{\omega}\cdot\mathbf{g}_i,
\end{align}
where $\hat{\omega}$ is a matrix whose elements are given by the corresponding components of the heterostrain tensor. The Dirac points are folded onto the corners of the new moir\'e Brillouin zone; the shift between them is\begin{align}
\label{eq:new_shift}
\mathbf{k}_0'=\boldsymbol{\kappa}_b'-\boldsymbol{\kappa}_t'\approx\boldsymbol{\kappa}_b-\boldsymbol{\kappa}_t+\hat{\omega}\cdot\frac{\mathbf{g}_1-\mathbf{g}_2}{3}.
\end{align}
This new shift modifies the moir\'e superlattice potential. When we write the Hamiltonian in the basis of Bloch states adapted to the new moir\'e lattice, the changes in the energetics of electrons do not enter through the interlayer matrix elements, second line of Eq.~\eqref{eq:matrix_Hamiltonian}, but through the new $\mathbf{G}$'s and $\boldsymbol{\kappa}_{\nu}$'s in the layer-diagonal terms. Hence, the \textit{disalignment} in the orientations between the beating pattern and the atomic lattices described in the main text enters in the Hamiltonain as a change of these vectors in the \textit{spinor} frame defined by $x$ and $y$ Pauli matrices acting on sublattice indices. In particular, treating the heterostrain-induced shift in the positions of the valleys (second term in Eq.~\ref{eq:new_shift}) in first order of perturbation theory leads to the energy shift of the Dirac points $\propto \alpha^2$ discussed in the main text. Interestingly, this geometrical energy shift is absent if the eigenstates labelled by different $\lambda$ become sublattice polarized in the limit $\beta=0$ (this was also checked numerically). This reflects the subtle interference phenomena behind the effect. The reason is that the additional chiral symmetry pins the Dirac points to zero energy. Nevertheless, the different strains accumulated on the two layers give rise to new couplings that break explicitly the chiral symmetry. For isotropic heterostrain, $\omega_{ij}=\bar{u}\,\delta_{ij}(=2u_{ij}^{t}=-2u_{ij}^{b})$, only the deformation potential matters,\begin{align}
\hat{\mathcal{V}}=\bar{u} \,D\, \hat{\nu}_z.
\end{align}
The band-structure calculations in Fig.~2 of the main text include both effects with a typical value of $D=10$ eV.

Uniform isotropic heterostrain considered so far describes the relative contraction of one layer with respect to the other, which could arise due to different couplings with the encapsulating boron nitrides. Deviations from this profile affect the excitonic insulator in different manners. Layer-asymmetric strain profiles are favored by the mutual Van der Waals interaction if the twist angle remains fixed. The predominance of layer-asymmetric perturbations is manifested, for example, in the evolution of the insulating states with displacement field. As discussed in the main text, heterostrain respects the approximate electron-hole symmetry of the bands, so it does not affect the condensation energy. Layer-symmetric strains, however, break electron-hole symmetry. Close to the magic angle, we can focus again on the deformation potential since the same interference process that suppresses $v_F^{*}$ makes the contribution from pseudo-gauge fields smaller. Space-dependent, layer-symmetric strain fields create diagonal disorder potentials $\hat{V}(\mathbf{r})=V(\mathbf{r})\hat{1}$, which have a pair-breaking effect akin to magnetic disorder in s-wave superconductors. The analogy between superconductors and excitonic insulators follows from the fact that time reversal symmetry in the former plays the same role as electron-hole symmetry in the latter.

\subsection{Long-range Coulomb interaction}

The long-range Coulomb interaction reads \begin{align}
H_{C}=\frac{1}{2}\int d\mathbf{r}\int d\mathbf{r}'\, V\left(\mathbf{r}-\mathbf{r}'\right) \hat{\rho}\left(\mathbf{r}\right)\hat{\rho}\left(\mathbf{r}'\right),
\end{align}
where $V(\mathbf{r})$ is the Coulomb potential and $\hat{\rho}(\mathbf{r})$ is the density operator. Within the continuum model, the latter is written as\begin{align}
\hat{\rho}\left(\mathbf{r}\right)=\sum_{\zeta,\alpha,\nu}\left[\hat{\psi}_{\zeta}^{\alpha,\nu}\left(\mathbf{r}\right)\right]^{\dagger}\hat{\psi}_{\zeta}^{\alpha,\nu}\left(\mathbf{r}\right).
\end{align}

It is convenient to introduce Fourier components of the density operators,\begin{subequations}\begin{align}
& \hat{\rho}\left(\mathbf{r}\right)=\frac{1}{\sqrt{A}}\sum_{\mathbf{k}} e^{i\mathbf{k}\cdot\mathbf{r}}\,\hat{\rho}\left(\mathbf{k}\right),\,\, \text{with}\\
& \hat{\rho}\left(\mathbf{k}\right)=\frac{1}{\sqrt{A}}\sum_{\mathbf{p}} \sum_{\zeta,\alpha,\nu}\left[\hat{c}_{\zeta}^{\alpha,\nu}\left(\mathbf{p}\right)\right]^{\dagger}\hat{c}_{\zeta}^{\alpha,\nu}\left(\mathbf{p}+\mathbf{k}\right).
\end{align}\end{subequations}
In these expressions, $\mathbf{k}$, $\mathbf{p}$ are not restricted to the moir\'e Brillouin zone, but are assumed to me smaller than the separation between microscopic valleys by construction. Introducing these series in the previous Hamiltonian yields to\begin{align}
H_{C}& =\frac{1}{2}\sum_{\mathbf{p}} V\left(\mathbf{p}\right)\,:\hat{\rho}\left(-\mathbf{p}\right)\hat{\rho}\left(\mathbf{p}\right):\\
&=\frac{1}{2A}\sum_{\mathbf{p},\mathbf{k}_1,\mathbf{k}_2} V\left(\mathbf{p}\right)\sum_{\alpha_1,\nu_1,\zeta_1}\sum_{\alpha_2,\nu_2,\zeta_2}\left[\hat{c}_{\zeta_1}^{\alpha_1,\nu_1}\left(\mathbf{k}_1+\mathbf{p}\right)\right]^{\dagger}\left[\hat{c}_{\zeta_2}^{\alpha_2,\nu_2}\left(\mathbf{k}_2-\mathbf{p}\right)\right]^{\dagger}\hat{c}_{\zeta_2}^{\alpha_2,\nu_2}\left(\mathbf{k}_2\right)\hat{c}_{\zeta_1}^{\alpha_1,\nu_1}\left(\mathbf{k}_1\right),
\nonumber
\end{align} 
where\begin{align}
V\left(\mathbf{p}\right)=\int d\mathbf{r}\,V\left(\mathbf{r}\right)e^{-i\mathbf{p}\cdot\mathbf{r}}=\frac{e^2\tanh\left(d|\mathbf{p}|\right)}{4\pi\epsilon\left|\mathbf{p}\right|}
\end{align}
for a double gate geometry. This Hamiltonian describes Coulomb scattering in a basis of plane waves corresponding to the first diagram in Fig.~3(a) of the main text. In this approximation we are already neglecting momentum exchange comparable with the separation between the microscopic valleys. This implies that rotational symmetry in the space of internal quantum number (e.g., spin) is preserved independently within each valley sector.

Next, we need to \textit{project} the interaction to the low-energy subspace. The projection procedure is really a truncation: the Coulomb interaction is expressed in the basis that diagonalizes the single-particle terms and then the summation in the new quantum numbers is limited to a few bands. By doing so we are neglecting interband coherences between the low-energy and remote bands. In principle, these should be included as exchange corrections to the low-energy band dispersion (through a renormalized group velocity $v_F^*$ in the low-energy theory).

The first step is to write momenta in components within the moir\'e Brillouin zone separated by vectors of the beating pattern,\begin{align}
\hat{\rho}\left(\mathbf{q}+\mathbf{G}\right)=\frac{1}{\sqrt{A}}\sum_{\mathbf{q}'\in\textrm{mBZ}}\sum_{\left\{\mathbf{G}'\right\}} \sum_{\zeta,\alpha,\nu}\left[\hat{c}_{\zeta}^{\alpha,\nu}\left(\mathbf{q}'+\mathbf{G}'\right)\right]^{\dagger}\hat{c}_{\zeta}^{\alpha,\nu}\left(\mathbf{q}+\mathbf{q}'+\mathbf{G}+\mathbf{G}'\right).
\end{align}
Using the relations in Eqs.~\eqref{eq:changing_basis_2nd}, we can rewrite this last expression as \begin{align}
\hat{\rho}\left(\mathbf{q}+\mathbf{G}\right)=\frac{1}{\sqrt{A}}\sum_{\mathbf{q}'\in\textrm{mBZ}}\sum_{\left\{\mathbf{G}'\right\}}\sum_{\lambda_1,\lambda_2} \sum_{\zeta,\alpha,\nu}\left[u_{\lambda_1,\zeta,\mathbf{G}'}^{\alpha,\nu}\left(\mathbf{q}'\right)\right]^{*}u_{\lambda_2,\zeta,\mathbf{G}+\mathbf{G}'}^{\alpha,\nu}\left(\mathbf{q}+\mathbf{q}'\right)\hat{c}_{\lambda_1,\zeta}^{\dagger}\left(\mathbf{q}'\right)\hat{c}_{\lambda_2,\zeta}\left(\mathbf{q}+\mathbf{q}'\right).
\end{align}
The notation is simplified if we introduce the following form factors,\begin{align}
\lambda_{\zeta,\mathbf{G}}^{(\lambda_1,\lambda_2)}\left(\mathbf{q}_1,\mathbf{q}_2\right)\equiv \sum_{\alpha,\nu}\sum_{\left\{\mathbf{G}'\right\}}\left[u_{\lambda_1,\zeta,\mathbf{G}'}^{\alpha,\nu}\left(\mathbf{q}_1\right)\right]^{*}u_{\lambda_2,\zeta,\mathbf{G}+\mathbf{G}'}^{\alpha,\nu}\left(\mathbf{q}_2\right)=\left\langle u_{\lambda_1,\zeta}\left(\mathbf{q}_1\right)\right|\hat{\mathcal{U}}_{\mathbf{G}}\left|u_{\lambda_2,\zeta}\left(\mathbf{q}_2\right) \right\rangle.
\end{align}
The density operator reads then\begin{align}
\hat{\rho}\left(\mathbf{q}+\mathbf{G}\right)=\frac{1}{\sqrt{A}}\sum_{\mathbf{q}'\in\textrm{mBZ}}\sum_{\lambda_1,\lambda_2}\sum_{\zeta} \lambda_{\zeta,\mathbf{G}}^{(\lambda_1,\lambda_2)}\left(\mathbf{q}',\mathbf{q}+\mathbf{q}'\right)\hat{c}_{\lambda_1,\zeta}^{\dagger}\left(\mathbf{q}'\right)\hat{c}_{\lambda_2,\zeta}\left(\mathbf{q}+\mathbf{q}'\right).
\end{align}
By plugging this expression into the Coulomb Hamiltonian,
\begin{align}
H_{C}=\frac{1}{2A}\sum_{\mathbf{q},\mathbf{q}_1,\mathbf{q}_2\in\textrm{mBZ}}\sum_{\lambda_1...\lambda_4} V_{\zeta_1,\zeta_2;\,\mathbf{q}_1,\mathbf{q}_2}^{\lambda_1,\lambda_2,\lambda_3,\lambda_4}\left(\mathbf{q}\right)\sum_{\lambda_1...\lambda_4}\hat{c}_{\lambda_1,\zeta_1}^{\dagger}\left(\mathbf{q}_1+\mathbf{q}\right)\hat{c}_{\lambda_3,\zeta_2}^{\dagger}\left(\mathbf{q}_2-\mathbf{q}\right)\hat{c}_{\lambda_4,\zeta_2}\left(\mathbf{q}_2\right)\hat{c}_{\lambda_2,\zeta_1}\left(\mathbf{q}_1\right),
\end{align}
and truncating the summation on band indices one obtains the projected interaction on a given subspace. The matrix elements read\begin{align}
V_{\zeta_1,\zeta_2;\,\mathbf{q}_1,\mathbf{q}_2}^{\lambda_1,\lambda_2,\lambda_3,\lambda_4}\left(\mathbf{q}\right)=\sum_{\left\{\mathbf{G}\right\}}V\left(\mathbf{q}+\mathbf{G}\right)\lambda_{\zeta_1,-\mathbf{G}}^{(\lambda_1,\lambda_2)}\left(\mathbf{q}+\mathbf{q}_1,\mathbf{q}_1\right)\lambda_{\zeta_2,\mathbf{G}}^{(\lambda_3,\lambda_4)}\left(\mathbf{q}_2-\mathbf{q},\mathbf{q}_2\right).
\end{align} 

The form factors describe variations of the charge density within the moir\'e cell and can be expressed as integrals of the form\begin{align}
\lambda_{\zeta,\mathbf{G}}^{(\lambda_1,\lambda_2)}\left(\mathbf{q}_1,\mathbf{q}_2\right)=\int d\mathbf{r}\, e^{-i\left(\mathbf{G}+\mathbf{q}_2-\mathbf{q}_1\right)\cdot\mathbf{r}}\left[u_{\lambda_1,\zeta,\mathbf{q}_1}\left(\mathbf{r}\right)\right]^*u_{\lambda_2,\zeta,\mathbf{q}_2}\left(\mathbf{r}\right),
\end{align}
which enter in the scattering amplitudes given in Eqs.~(7) of the main text. To arrive at those expressions, note that in the low-energy theory we are only concerned about scattering events between electronic quasiparticles around $\boldsymbol{\kappa}_{\nu}$ points in the lowest-energy bands. The summations on incoming momenta $\mathbf{q}_{1,2}$ are then restricted to small deviations away from these points, $\mathbf{q}_i=\boldsymbol{\kappa}_{\nu}+\boldsymbol{p}_i$:\begin{align}
\sum_{\mathbf{q}_i\in\textrm{mBZ}}\longrightarrow\sum_{\{\boldsymbol{\kappa}_\nu\}}\sum_{\boldsymbol{p}_i}
\end{align}
We can apply the same philosophy to the exchanged momentum $\mathbf{q}$; we may consider two scenarios, either $\mathbf{q}$ is small compared with the separation between mini-valleys, $\mathbf{k}_0$, or it is comparable and the electronic state is scattered to a different mini-valley. These two types of processes (\textit{intra-} and \textit{inter-node} scattering) are represented by the two interaction vertices in Fig.~3(a) of the main text labelled by $\eta=0$ and $\eta=\pm 1$, respectively. Their matrix elements in the 8-spinor basis read\begin{subequations}
\label{eq:Coulomb_matrix_elements}
\begin{align}
& V_{\zeta_1,\zeta_2;\,\boldsymbol{\kappa}_{\nu},\boldsymbol{\kappa}_{\nu}}^{\lambda_1,\lambda_2,\lambda_3,\lambda_4}\left(\boldsymbol{p}\right)=\sum_{\left\{\mathbf{G}\right\}}V\left(\boldsymbol{p}+\mathbf{G}\right)\,\lambda_{\zeta_1,-\mathbf{G}}^{(\lambda_1,\lambda_2)}\left(\boldsymbol{\kappa}_{\nu},\boldsymbol{\kappa}_{\nu}\right) \lambda_{\zeta_2,\mathbf{G}}^{(\lambda_3,\lambda_4)}\left(\boldsymbol{\kappa}_{\nu},\boldsymbol{\kappa}_{\nu}\right),\\
& V_{\zeta_1,\zeta_2;\,\eta=+1}^{\lambda_1,\lambda_2,\lambda_3,\lambda_4}\left(\boldsymbol{p}\right)=\sum_{\left\{\mathbf{G}\right\}}V\left(\boldsymbol{p}+\mathbf{k}_0+\mathbf{G}\right)\,\lambda_{\zeta_1,-\mathbf{G}}^{(\lambda_1,\lambda_2)}\left(\boldsymbol{\kappa}_{b},\boldsymbol{\kappa}_{t}\right)\lambda_{\zeta_2,\mathbf{G}}^{(\lambda_3,\lambda_4)}\left(\boldsymbol{\kappa}_{t},\boldsymbol{\kappa}_{b}\right),\\
& V_{\zeta_1,\zeta_2;\,\eta=-1}^{\lambda_1,\lambda_2,\lambda_3,\lambda_4}\left(\boldsymbol{p}\right)= \sum_{\left\{\mathbf{G}\right\}}V\left(\boldsymbol{p}-\mathbf{k}_0+\mathbf{G}\right)\,\lambda_{\zeta_1,-\mathbf{G}}^{(\lambda_1,\lambda_2)}\left(\boldsymbol{\kappa}_{t},\boldsymbol{\kappa}_{b}\right)\lambda_{\zeta_2,\mathbf{G}}^{(\lambda_3,\lambda_4)}\left(\boldsymbol{\kappa}_{b},\boldsymbol{\kappa}_{t}\right).
\end{align}
\end{subequations}

Finally, in the BCS-like mean-field calculation of the main text we are only retaining scattering events in the dominant electron-hole channel. Other processes enter in the band-diagonal terms of the Hartree-Fock self-energy, which are assumed to be included in the single-particle energies. Among the dominant terms, I argued that inter-node matrix elements are smaller and contribute less to build the amplitude of the excitonic condensate than small-momenta intra-node scattering. Nevertheless, it is worth emphasizing that these processes play an important role in defining the matrix structure of the order parameter. In the absence of inter-node Coulomb scattering, the effective low-energy Hamiltonian would remain invariant under an enlarged continuous group formed by independent valley rotations within each mini-valley $\boldsymbol{\kappa}_{\nu}$. In that case, the order-parameter manifold would be U(1)$\times$U(1), parametrized by two independent phases. Momentum exchange between mini-valleys is ultimately responsible for locking the relative phase of electron-hole pairing on each $\boldsymbol{\kappa}_{\nu}$ sector.

\section{Bound states in vortex excitations}

Consider the mean-field Hamiltonian $\hat{\mathcal{H}}_0+\hat{\Delta}$. The excitonic condensate breaks $\mathcal{T}$ and U$_{v}$(1) symmetries, but preserves the combination of $\mathcal{T}$ and valley rotations $e^{i\frac{\pi}{2}\hat{\Lambda}_z}=i\hat{\Lambda}_z$:\begin{align}
\mathcal{T}':\,\hat{\Sigma}_y\hat{\Lambda}_x\hat{\Gamma}_x\mathcal{K}.
\end{align}
This is a Kramers ($\mathcal{T}'\,^2=-1$) time-reversal operation associated with the pseudo-spin degree of freedom. Additionally, if we only keep the Dirac dispersion, the Hamiltonian respects an emergent particle-hole symmetry given by\begin{align}
\Theta:\, i\,\hat{\Sigma}_x\hat{\Lambda}_x\mathcal{K}.
\end{align}
The effective mean-field Hamiltonian belongs to class DIII, whose point defects are characterized by a $\mathbb{Z}_2$ index. In fact, the Hamiltonian can be seen as two copies of the Jackiw-Rossi model on each mini-valley sector connected by $\mathcal{T}'$ symmetry. In that model, textures with vorticity $n$ host $|n|$ zero modes localized in their core. In the present case, the number of zero modes is double, $2|n|$, due to Kramers degeneracy. At least a pair of Kramers partners remains pinned to zero energy in the presence of $\Theta$ symmetry if $n_{\textrm{mod}2}=1$.

Let us focus on vortex excitations in the orbital sector with $|n|=1$, as described in the main text. The effective Hamiltonian for such a vortex configuration reads\begin{align}
\label{eq:H_MF}
\hat{\mathcal{H}}\left(\mathbf{r}\right)=-i\hbar v_F^*\, \hat{\boldsymbol{\Sigma}}\cdot\boldsymbol{\partial}+2 g_{A_2}\bar{u}\,\hat{\Lambda}_z\hat{\Gamma}_z+\Delta_0\left(r\right)\left[\cos\vartheta\,\hat{\Sigma}_z\hat{\Lambda}_x\hat{\Gamma}_z\pm\sin\vartheta\, \hat{\Sigma}_z\hat{\Lambda}_y\hat{\Gamma}_z\right],
\end{align}
where positions are expressed in polar coordinates $\mathbf{r}=(r,\vartheta)$ referred to the vortex core, and $\Delta(r)>0$ everywhere except at the origin, $\Delta_0(r\rightarrow 0)\rightarrow 0$; the upper/lower sign in the last term corresponds to vorticity $n=\pm 1$. The Hamiltonian can be diagonalized in the basis of eigenvectors of the generalized angular momentum operator\begin{align}
\hat{J}=-i\partial_{\vartheta}+\frac{1}{2}\hat{\Sigma}_z\pm\frac{1}{2}\hat{\Lambda}_z,
\end{align}
which reads (note that $n=\pm1$)\begin{align}
\boldsymbol{\tilde{u}}_{m}\left(r,\vartheta\right)=e^{im\vartheta}\left[\begin{array}{c}
e^{-\frac{i(n+1)\vartheta}{2}}\zeta_1(r)\\
i\,e^{-\frac{i(n-1)\vartheta}{2}}\zeta_2(r)\\
e^{\frac{i(n-1)\vartheta}{2}}\zeta_3(r)\\
i\,e^{\frac{i(n+1)\vartheta}{2}}\zeta_4(r)\\
e^{-\frac{i(n+1)\vartheta}{2}}\eta_1(r)\\
i\,e^{-\frac{i(n-1)\vartheta}{2}}\eta_2(r)\\
e^{\frac{i(n-1)\vartheta}{2}}\eta_3(r)\\
i\,e^{\frac{i(n+1)\vartheta}{2}}\eta_4(r)
\end{array}\right],
\end{align}
where $\zeta_i(r)$, $\eta_i(r)$ can be taken to be real. Note that $\zeta_i(r)$ and $\eta_i(r)$ are not mixed. $\mathcal{T}'$ symmetry connects solutions with opposite angular momentum $m$ such that $\eta_1(r)=-\zeta_4(r)$, $\eta_2(r)=\zeta_3(r)$, $\eta_3(r)=-\zeta_2(r)$, and $\eta_4(r)=\zeta_1(r)$. Topologically protected zero modes can only exist in the $m=0$ channel, otherwise there would be an extra orbital degeneracy and the modes would be gapped by generic perturbations. For zero-energy modes, $\Theta$ symmetry implies that $\zeta_1(r)=\varpi\,\zeta_4(r)$, $\zeta_2(r)=\varpi\,\zeta_3(r)$, with $\varpi=\pm 1$ as $\Theta^2=1$.

Consider first the case $n=+1$, $\varpi=+1$. Zero-energy solutions satisfy \begin{subequations}\begin{align}
& 2g_{A_2}\bar{u}\,\zeta_1(r)+\left[\hbar v_F^*\partial_r+\Delta_0(r)\right]\zeta_2(r)=0,\\
& \left[\hbar v_F^*\left(\partial_r+\frac{1}{r}\right)+\Delta_0(r)\right]\zeta_1(r)-2g_{A_2}\bar{u}\,\zeta_2(r)=0.
\end{align}
\end{subequations}
The solutions to this set of equations are\begin{subequations}\begin{align}
& \zeta_1(r)=e^{-\frac{1}{\hbar v_F^*}\int_0^{r}dr'\,\Delta_0(r')}J_1\left(\frac{2g_{A_2}\bar{u}}{\hbar v_F^*}\, r\right),\\
& \zeta_2(r)=e^{-\frac{1}{\hbar v_F^*}\int_0^{r}dr'\,\Delta_0(r')}J_0\left(\frac{2g_{A_2}\bar{u}}{\hbar v_F^*}\, r\right),
\end{align}
\end{subequations}
where $J_{0,1}(x)$ are Bessel functions of the first kind. In the case of $\varpi=-1$ the solutions are not normalizable. In the case of vorticity $n=-1$, it is the other way around, normalizable solution are those associated with $\varpi=-1$.

It follows then that textures with vorticity $\pm 1$ host a pair of bound states of zero energy connected by $\mathcal{T}'$ symmetry, as expected from the previous arguments. Particle-hole symmetry also requieres that their spectral weight must be equally borrowed from the positive and negative energy continuum. Accounting for the additional spin degeneracy, the core of a vortex hosts four bound states, so the presence of one of these textures creates a defect of $2e$ charge around it. Vortex/anti-vortex are created in pairs, so there are four electrons that must be accommodated in eight available bound states. The spin $S$ and charge $Q$ quantum numbers of vortices depend on the particular arrangement; for example, a triply occupied vortex carries $Q=-e$, $S=\pm 1/2$, and the single occupied anti-vortex the opposite numbers, $Q=e$, $S=\mp 1/2$, and so on. These arguments are similar to the case of Kekul\'e bond order in graphene, however, in that case there is spin-charge separation (charge fractionalization masked by the spin degeneracy), which is absent in the present case due to the additional Kramers degeneracy in the orbital sector. 

\section{Landau level degeneracy}

Consider the mean-field Hamiltonian in Eq.~\eqref{eq:H_MF} for a uniform order parameter in Peierls substitution, $-i\boldsymbol{\partial}\rightarrow\hat{\boldsymbol{\pi}}\equiv-i\boldsymbol{\partial}+e\mathbf{A}/c$, where $B=\boldsymbol{\nabla}\times\mathbf{A}|_z$ is an external out-of-plane magnetic field. In matrix notation, the Hamiltonian reads\begin{align}
\hat{\mathcal{H}}_{\pm}=\left[\begin{array}{cccc}
\pm2g_{A_2}\tilde{u} & \hbar\omega_c\hat{a} & \mp i\Delta_0 & 0\\
\hbar\omega_c\hat{a}^{\dagger} & \mp2g_{A_2}\tilde{u} & 0 &  \mp i\Delta_0\\
 \pm i\Delta_0 & 0 & \pm2g_{A_2}\tilde{u} & \hbar\omega_c\hat{a}\\
 0 & \mp i\Delta_0 & \hbar\omega_c\hat{a}^{\dagger} & \pm 2g_{A_2}\tilde{u}
\end{array}\right],
\end{align}
where $\omega_c=\sqrt{2}v_F^{*}\ell_B$ is the cyclotron frequency with magnetic length $\ell_B=\sqrt{c/eB}$, $\hat{a}\equiv(\pi_x-i\pi_y)\ell_B/\sqrt{2}$, and the upper (lower) sign applies to mini-valley $\boldsymbol{\kappa}_t$ ($\boldsymbol{\kappa}_{b}$).

In the Landau gauge $\mathbf{A}=(-By,0)$, $\hat{a}$/$\hat{a}^{\dagger}$ act as lowering/raising operators in the basis of functions $e^{ikx}\phi_n(y)$, where $\phi_n(y)$ with $n\geq 1$ are solutions of the 1D oscillator in the coordinate $y/\ell_B-\ell_B k$. For $n>0$, solutions of the previous Hamiltonian are of the form\begin{align}
\boldsymbol{\tilde{u}}_{n}\left(x,y\right)=e^{ikx}\left[\begin{array}{c}
c_1\,\phi_{n-1}(y)\\
c_2\,\phi_{n}(y)\\
c_3\,\phi_{n-1}(y)\\
c_4\,\phi_{n}(y)
\end{array}\right],
\end{align}
where the coefficients $c_{i}$ as well as the eigenenergies $\varepsilon_n$ follow from the solution of the following secular equation:\begin{align}
\left[\begin{array}{cccc}
\pm2g_{A_2}\tilde{u} & \hbar\omega_c\sqrt{n} & \mp i\Delta_0 & 0\\
\hbar\omega_c\sqrt{n} & \mp2g_{A_2}\tilde{u} & 0 &  \mp i\Delta_0\\
 \pm i\Delta_0 & 0 & \pm2g_{A_2}\tilde{u} & \hbar\omega_c\sqrt{n}\\
 0 & \mp i\Delta_0 & \hbar\omega_c\sqrt{n} & \pm 2g_{A_2}\tilde{u}
\end{array}\right]\cdot \left[\begin{array}{c}
c_1\\
c_2\\
c_3\\
c_4
\end{array}\right]=\varepsilon_n\left[\begin{array}{c}
c_1\\
c_2\\
c_3\\
c_4
\end{array}\right].
\end{align}
The energies $n\neq 0$ Landau levels are degenerate in mini-valley and given by\begin{align}
\varepsilon_n=\pm\sqrt{\Delta_0^2+\left(\sqrt{n}\hbar\omega_c\pm2 g_{A_2}\tilde{u}\right)^2}.
\end{align}
For $n=0$, the previous ansatz only makes sense for $c_1=c_3=0$. The secular equation reduces to \begin{align}
\left[\begin{array}{cc}
\pm2g_{A_2}\tilde{u} & \pm i\Delta_0\\
\mp i\Delta_0 & \mp g_{A_2}\tilde{u}
\end{array}\right]\cdot \left[\begin{array}{c}
c_2\\
c_4
\end{array}\right]=\varepsilon_0\left[\begin{array}{c}
c_2\\
c_4
\end{array}\right],
\end{align}
and the zero-Landau level energies reduces to $\pm\sqrt{\Delta_0^2+4\, g_{A_2}^2\tilde{u}^2}$, as prescribed by the previous sequence.

Accounting for spin, the Landau levels are 4-fold degenerate. If, additionally, C$_{2z}$ symmetry is broken (either spontaneously or by a residual coupling with the encapsulating boron nitrides), so a staggered potential of the form $M\,\Sigma_z\Lambda_z$ is also present, then the sequence of Landau levels reads\begin{align}
\varepsilon_n=\pm\sqrt{\Delta_0^2+\left(\sqrt{n\hbar^2\omega_c^2+M^2}\pm2 g_{A_2}\tilde{u}\right)^2}.
\end{align}
Thus, the orbital degeneracy of the $n=0$ Landau level is removed. According to this $k\cdot p$ model, the Landau level sequence in the Landau fan diagram emanating from neutrality would be $\pm 2,\pm4,\pm8,\pm12...$, which corresponds to the sequence observed in the insulating devices of Ref.~9.

\end{document}